\documentclass[prl,twocolumn,notitlepage,longbibliography]{revtex4-2}

\usepackage[utf8]{inputenc}
\usepackage{mathrsfs}
\usepackage{graphicx}
\usepackage{amsmath}
\usepackage{amssymb}
\usepackage{amsfonts}
\usepackage{color}
\usepackage[unicode=true,colorlinks=true,citecolor=blue,urlcolor=blue]{hyperref}
\usepackage[normalem]{ulem}
\usepackage{epsfig}
\usepackage{bm}

\DeclareUnicodeCharacter{2212}{-}

\begin{document}
	
\title{Robust Interaction-Enhanced Sensing via Antisymmetric Rabi Spectroscopy}

\author{Jiahao Huang$^{1,2,3}$}
\altaffiliation{Email: hjiahao@mail2.sysu.edu.cn, eqjiahao@gmail.com}

\author{Sijie Chen$^{2,3}$}

\author{Min Zhuang$^{1}$}

\author{Chaohong Lee$^{1,2,3}$}
\altaffiliation{Email: chleecn@szu.edu.cn, chleecn@gmail.com}

\affiliation{$^{1}$College of Physics and Optoelectronic Engineering, Shenzhen University, Shenzhen 518060, China}
\affiliation{$^{2}$Guangdong Provincial Key Laboratory of Quantum Engineering and Quantum Metrology, School of Physics and Astronomy, Sun Yat-Sen University (Zhuhai Campus), Zhuhai 519082, China}
\affiliation{$^{3}$State Key Laboratory of Optoelectronic Materials and Technologies, Sun Yat-Sen University (Guangzhou Campus), Guangzhou 510275, China}
	
\begin{abstract}
Atomic spectroscopy, an essential tool for frequency estimation, is widely used in quantum sensing.
Atom-atom interaction can be used to generate entanglement for achieving quantum enhanced sensing.
However, atom-atom interaction always induces collision shift, which brings systematic error in determining the resonance frequency.
Contradiction between utilizing atom-atom interaction and suppressing collision shift generally exists in atomic spectroscopy.
Here, we propose an antisymmetric Rabi spectroscopy protocol without collision shift in the presence of atom-atom interactions.
We analytically find that the antisymmetric point can be used for determining the resonance frequency.
For small Rabi frequency, our antisymmetric Rabi spectroscopy with slight atom-atom interaction can provide better measurement precision than the conventional Rabi spectroscopy.
With stronger atom-atom interaction and Rabi frequency, the spectrum resolution can be dramatically improved and the measurement precision may even beat the standard quantum limit.
Moreover, unlike the quantum-enhanced Ramsey interferometry via spin squeezing, our scheme is robust against detection noises.
Our antisymmetric Rabi spectroscopy protocol has promising applications in various practical quantum sensors such as atomic clocks and atomic magnetometers.

\end{abstract}	

\date{\today}

\maketitle

\textit{Introduction. --}
Atomic spectroscopy, the basis for precise measurement of transition frequencies, has been extensively applied in fundamental sciences as well as high-precision sensors~\cite{Degen2017}.
Most quantum sensors employ a pair of discrete energy levels relevant to the physical quantity to be measured.
By using atomic spectroscopy to determine the transition frequency, one can estimate the physical quantity to be measured~\cite{5778937} and build various practical quantum sensors such as atomic clocks~\cite{Diddams2004,Oelker2019,Orenes2021}, magnetometers~\cite{Baumgart2016,PhysRevA.80.033420,Hao2018}, gyroscopes~\cite{4319483} and gravimeters~\cite{Szigeti2021}.
On one hand, atom-atom interaction can be used to generate the desired entanglement for improving the measurement precision from the standard quantum limit (SQL) to the Heisenberg limit (HL)~\cite{Giovannetti2006, Lee2006, Huang2014, Pezze2018, Lu2019}.
On the other hand, in atomic spectroscopy~\cite{Rabi1938,Ramsey1949,Ramsey1950,Ramsey1980}, atom-atom interaction always results in collision shift~\cite{Ludlow2015} which brings systematic error in determining the resonance frequency~\cite{PhysRevA.56.R4389}.
Many schemes have been developed to suppress the influences of atom-atom collisions~\cite{Szymaniec2007,PhysRevLett.104.010801,Lee2016}.

Generally, in high-precision quantum sensing via atomic spectroscopy, there is a contradiction between utilizing atom-atom interaction and suppressing collision shift.
To achieve high-precision Ramsey spectroscopy, one can use atom-atom interaction to prepare the desired entanglement and then turn off atom-atom interaction for interrogation.
In this way, the collision shift can be small, but it requires precise time-dependent manipulation of atom-atom interactions.
In Rabi spectroscopy, the coupling field and the atom-atom interaction coexist.
It has been demonstrated that one may achieve a Rabi spectroscopy towards the HL with two correlated ions~\cite{PhysRevLett.120.243603}.
However, in such a protocol, the system should be split into two orthogonal subspaces with different parities as different probes, which adds additional complexity to the existing Rabi protocols.
Therefore, it starves for an efficient spectroscopy protocol without collision shift and meanwhile one can utilize atom-atom interactions to improve the measurement precision beyond the SQL.

In this Letter, we propose a robust interaction-enhanced frequency estimation protocol via antisymmetric Rabi spectroscopy.
Instead of preparing all atoms in the lower level for the conventional Rabi spectroscopy, our protocol inputs an equal superposition state of the two sensor levels before applying the coupling field for implementing Rabi oscillation.
We analytically find that the Rabi spectrum becomes exactly antisymmetric with respect to the detuning in the presence of atom-atom interactions.
This means the absence of collision shift in our protocol theoretically.
In practice, the desired input state can be generated by applying a short $\pi/2$ pulse and the antisymmetric point can still appear even under atom-atom interaction.
For small Rabi frequency, our protocol with slight atom-atom interaction can offer better measurement precision.
More importantly, the measurement precision can be improved beyond the SQL by using stronger atom-atom interaction and Rabi frequency.
Compared to the spin-squeezing-enhanced Ramsey spectroscopy protocol, our scheme is much more robust against detection noises.

\begin{figure}[htb]
\centering
\includegraphics[width=\columnwidth]{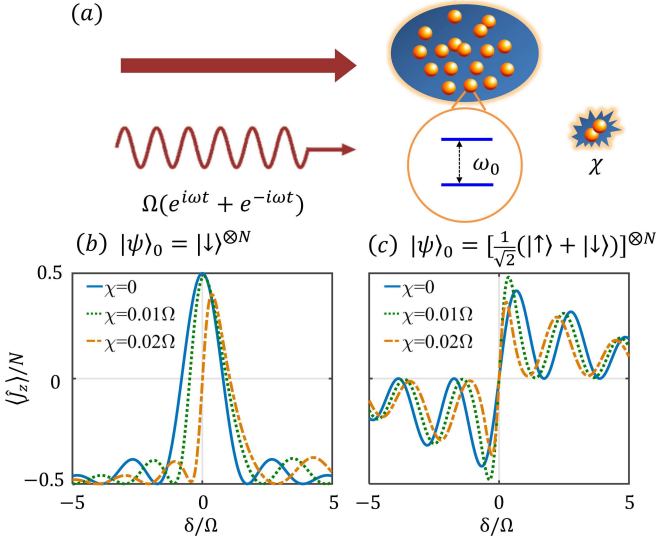}\caption{(Color online) (a) Schematic of a quantum sensor via Rabi oscillation. A laser field interacts with an ensemble of interacting atoms. $\Omega$ denotes the Rabi frequency, $\omega$ is the laser frequency, $\omega_0$ is the atomic transition frequency to be measured, and $\chi$ denotes the atom-atom interaction. (b) The final population difference versus the detuning $\delta$ for the initial state $|\psi\rangle_0=|\pi,0\rangle_{SCS}$. (c) The final population difference versus the detuning $\delta$ for the initial state $|\psi\rangle_0=|\pi/2,0\rangle_{SCS}$, which can be realized by applying a short $\pi/2$ pulse along $y$ axis on $|\pi,0\rangle_{SCS}$. Here, the atom number $N=100$ and the total evolution time $T$ satisfying $\Omega T=\pi$. The blue (solid), green (dashed) and orange (dash-dotted) lines are the results with $\chi=0$, $0.01\Omega$ and $0.02\Omega$, respectively.}
\label{Fig1}
\end{figure}

\textit{Conventional Rabi oscillation in an atomic ensemble. --}
We consider an ensemble of $N$ two-level atoms, which can be regraded as identical pseudospin-$\frac{1}{2}$ particles obeying a collective spin $\hat{J}$ with $\hat{J}_x = \sum_{n=1}^{N} \sigma_x^{(n)}$, $\hat{J}_y = \sum_{n=1}^{N} \sigma_y^{(n)}$, and $\hat{J}_z = \sum_{n=1}^{N} \sigma_z^{(n)}$.
Here $\sigma_{x,y,z}^{(n)}$ denotes the Pauli matrices for the $n$-th atom.
In the Schwinger representation, $\hat{J_x}=\frac {1}{2}(\hat{a}^{\dag}\hat{b}+\hat{b}^{\dag}\hat{a})$, $\hat{J_y}=\frac {i}{2}(\hat{a}^{\dag}\hat{b}-\hat{b}^{\dag}\hat{a})$, $\hat{J_z}=\frac {1}{2}(\hat{b}^{\dag}\hat{b}-\hat{a}^{\dag}\hat{a})$ , where $\hat{a}$ and $\hat{b}$ are the annihilation operators for atoms in $\left|\uparrow\right\rangle$ and $\left|\downarrow\right\rangle$, respectively.
These collective spin operators obey the general angular momentum commutation relations $[\hat{J_i},\hat{J_j}]=i\hbar\epsilon_{ijk}\hat{J_k}$ with $i,j,k=x,y,z$ and $\epsilon_{ijk}$ the Levi-Civita symbol.
Thus, an arbitrary state can be expressed by $|\Psi\rangle=\sum_{m=-N/2}^{N/2} C_{m}|J, m\rangle$, where $|J, m\rangle$ is the Dicke basis denoting $(N/2-m)$ atoms in $\left|\uparrow\right\rangle$ and $(N/2+m)$ atoms in $\left|\downarrow\right\rangle$.

A well-known spectroscopic method for frequency estimation is implementing Rabi oscillations.
In a non-interacting atomic ensemble driven by an external coupling field, its Rabi oscillations obey the Hamiltonian $\hat H=\omega_0 \hat{J_z} + \Omega (e^{i\omega t}+e^{-i\omega t}) \hat{J_x}$, where $\omega_0$ is the atomic transition frequency, $\omega$ is the frequency of the coupling field, and $\Omega$ is the Rabi frequency. For simplicity we set $\hbar=1$ hereafter.
In the rotating-frame with rotating-wave approximation (RWA), the Hamiltonian becomes
\begin{equation}\label{Ham_rabi}
	\hat H_R=\Omega \hat{J_x} + \delta \hat{J_z},
\end{equation}
where the detuning $\delta=\omega_0-\omega$.
Without atom-atom interactions, the system state can be described by a spin coherent state (SCS) $\left|\theta,\phi\right\rangle_{SCS}=\left(\cos \theta/2 \left|\uparrow\right\rangle +e^{i\phi}\sin \theta/2 \left|\downarrow\right\rangle\right)^{\otimes N}$, in which all atoms are in the same quantum state.
Conventionally, one prepare an initial SCS $\left|\psi\right\rangle_0 = \left|\pi,0\right\rangle_{SCS}=\left(\left|\downarrow\right\rangle\right)^{\otimes N}$ with all atoms in $\left|\downarrow\right\rangle$.
Then, the initial state evolves according to $\left|\psi(t)\right\rangle = e^{-i \hat H_R t}\left|\pi,0\right\rangle_{SCS}$.
%
%
At time $T$, we have the half population difference
\begin{eqnarray}\label{Jz}
	\langle\hat{J_z}(T)\rangle=-\frac{N}{2}\frac{\delta^2+ \Omega^2 \cos (\sqrt{\Omega^2+\delta^2}T)}{\Omega^2+\delta^2},
\end{eqnarray}
which is a symmetric function with respect to $\delta=0$.
%
%
In practice, one would choose $\Omega T=\pi$ and measure the half population difference $\langle\hat{J_z}(T)\rangle$.
It attains the maximum at $\delta=0$ so that can be used as the frequency locking signal, see the blue solid line in Fig.~\ref{Fig1}~(b).

Taking into account the atom-atom interaction, the original Hamiltonian becomes $\hat H'=\omega_0 \hat{J_z} + \chi \hat{J}_z^2+ \Omega (e^{i\omega t}+e^{-i\omega t}) \hat{J_x}$, where $\chi$ characterizes the strength of effective atom-atom interaction [see Fig.~\ref{Fig1}~(a)].
In the rotating-frame with RWA, the Hamiltonian reads
\begin{equation}\label{Ham_rabi2}
	\hat H_R'=\Omega \hat{J_x} + \chi \hat{J}_z^2 + \delta \hat{J_z}.
\end{equation}
When $\chi$ is non-negligible, the atomic collision causes a frequency shift and the locking signal is no longer symmetric with $\delta=0$.
For example, when $\chi=0.01\Omega$, an obvious frequency shift can be observed, see the green dashed line in Fig.~\ref{Fig1}~(b).
While for $\chi=0.02\Omega$, the frequency shift is larger and the lineshape also distorted, see the orange dash-dotted line in Fig.~\ref{Fig1}~(b).

In the conventional Rabi spectroscopy, the atom-atom interaction always induces a collision shift that decreasing the accuracy for estimating the transition frequency.
Besides, since the maximum is around $\delta=0$, its signal slope equals zero and the standard deviation $\Delta \delta$ at this point is diverged.
In general, it is necessary to find out two symmetric points (with respect to $\delta=0$) to determine the on-resonance point.
These limit the performances of the Rabi spectroscopy.
Below, we analytically analyze the quantum sensing via antisymmetric Rabi spectroscopy and demonstrate how to achieve better frequency estimation at the on-resonance point $\delta=0$.

\begin{figure}[htb]
\centering
\includegraphics[width=\columnwidth]{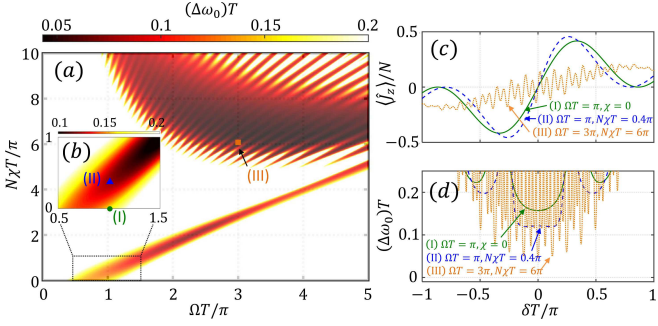}\caption{(Color online) (a) The measurement precision $\Delta\omega_0$ at the on-resonance point versus the Rabi frequency $\Omega$ and the atom-atom interaction $\chi$ for a fixed evolution time $T=1$. (b) The enlarged area of the dotted box in (a). (c) The scaled population difference $\langle\hat{J_z}(T)\rangle/N$ and (d) the measurement precision $\Delta \omega_0$ versus the detuning $\delta$ for the points I (green circle, $\Omega=\pi, \chi=0$), II (blue triangle, $\Omega=\pi, \chi=0.4\pi/N$) and III (orange square, $\Omega=3\pi, \chi=6\pi/N$). Here, the total atom number $N=100$ and the initial state $|\psi\rangle_0=|\pi/2,0\rangle_{SCS}$. }
\label{Fig2}
\end{figure}

\textit{Antisymmetric Rabi spectroscopy. --}
Let's turn back to the non-interacting case ($\chi=0$).
For the Rabi oscillation from an arbitrary initial state $|\psi\rangle_0$, we analytically obtain its half population difference $\langle \hat J_z(T)\rangle$ at time $T$,
\begin{eqnarray}\label{ana1}
&&\langle \hat J_z(T)\rangle=\langle \psi(T)|\hat{J_z}|\psi(T)\rangle = _0\!\!\langle \psi| e^{i \hat H_R T} \hat{J_z} e^{-i \hat H_R T}|\psi\rangle_0 \nonumber\\
&& =\langle\hat{J_x}\rangle_0 \left(\sum_{k=1}^{\infty}\frac{1}{(2k)!} x_{2k}\right)+\langle\hat{J_y}\rangle_0 \left(\sum_{k=1}^{\infty}\frac{1}{(2k-1)!} y_{2k-1}\right)\nonumber\\
&& ~~+\langle\hat{J_z}\rangle_0\left(1+\sum_{k=1}^{\infty}\frac{1}{(2k)!} z_{2k}\right) \nonumber\\
&&=\langle\hat{J_x}\rangle_0 X(\Omega,\delta,T)\!+\!\langle\hat{J_y}\rangle_0 Y(\Omega,\delta,T)\!+\! \langle\hat{J_z}\rangle_0 Z(\Omega,\delta,T),
\end{eqnarray}
where $y_1=\Omega T$, $x_{2k}=y_{2k-1} (\delta T)$, $z_{2k}=-y_{2k-1} (\Omega T)$ and $y_{2k-1}=-(\delta T)x_{2k} +(\Omega T)z_{2k}$ for $k>1$.
The sums $X(\Omega,\delta,T)=\frac{\Omega\delta}{\Omega^2+\delta^2}\left[1-\cos (\sqrt{\Omega^2+\delta^2} T)\right]$, $Y(\Omega,\delta,T)=\frac{\Omega \sin (\sqrt{\Omega^2+\delta^2} T)}{\sqrt{\Omega^2+\delta^2}}$, and $Z(\Omega,\delta,T)=\frac{\delta^2+ \Omega^2 \cos (\sqrt{\Omega^2+\delta^2} T)}{\Omega^2+\delta^2}$ can be analytically calculated, see more details~\cite{SM}.
If $|\psi\rangle_0=|\pi,0\rangle_{SCS}$, we have $\langle\hat{J_z}\rangle_0=-\frac{N}{2}$ and $\langle\hat{J_x}\rangle_0=\langle\hat{J_y}\rangle_0=0$ and Eq.~\eqref{ana1} gives Eq.~\eqref{Jz}.

According to Eq.~\eqref{ana1}, one can analytically give the relation between the half population difference $\langle \hat J_z(T)\rangle$ and the detuning $\delta$ for any initial state $|\psi\rangle_0$.
Since $X(\Omega,\delta,t)=-X(\Omega,-\delta,t)$, $Y(\Omega,\delta,t)=Y(\Omega,-\delta,t)$ and $Z(\Omega,\delta,t)=Z(\Omega,-\delta,t)$, we find that, when $\langle\hat{J_z}\rangle_0=\langle\hat{J_y}\rangle_0=0$, the half population difference $\langle \hat J_z(T)\rangle=\frac{ \Omega\delta}{\Omega^2+\delta^2}\left[1-\cos (\sqrt{\Omega^2+\delta^2} T)\right]\langle\hat{J_x}\rangle_0$ is exactly antisymmetric with respect to $\delta=0$.
For an example, inputting the initial state $|\psi\rangle_0=|\pi/2,0\rangle_{SCS}$ whose $\langle\hat{J_x}\rangle_0=\frac{N}{2}$ and $\langle\hat{J_z}\rangle_0=\langle\hat{J_y}\rangle_0=0$, we have
\begin{equation}\label{Jz_SCSx}
    \langle \hat J_z(T)\rangle= \frac{N}{2} \frac{\Omega\delta}{\Omega^2+\delta^2}\left[1-\cos (\sqrt{\Omega^2+\delta^2} T)\right].
\end{equation}
Eq.~\eqref{Jz_SCSx} is an antisymmetric function of $\delta$ for arbitrary $\Omega$ and $T$, see the blue solid line in Fig.~\ref{Fig1}~(c).
The measurement precision can be calculated as
\begin{equation}\label{precision}
	\Delta\omega_0=\frac{\Delta \hat{J}_z}{|\partial \langle\hat{J}_z\rangle/\partial \omega_0|}=\frac{\Delta \hat{J}_z}{|\partial \langle\hat{J}_z\rangle/\partial \delta|}.
\end{equation}
Compared to the conventional Rabi spectroscopy with $|\psi\rangle_0=|\pi,0\rangle_{SCS}$, the slope of the signal at on-resonance point $\delta=0$ becomes sharp and the corresponding measurement precision becomes high~\cite{Ramsey1980,Sanner2018}.

Amazingly, in the presence of atom-atom interaction, there is no collision shift in the antisymmetric Rabi spectroscopy.
We find that the frequency locking signal is always antisymmetric with respect to $\delta=0$ in the presence of $\chi$, see Fig.~\ref{Fig1}~(c).
Since the antisymmetry of the signal preserves with $\chi$, the antisymmetric point has no collision shift theoretically and can be used as the on-resonance point.

The key for antisymmetric Rabi spectroscopy originates the symmetry of Hamiltonian~\eqref{Ham_rabi2}~\cite{PhysRevLett.102.070401,Trenkwalder2016,Zhuang2020}.
Under the transformation $\hat a (\hat b) \rightarrow \hat b (\hat a)$, we have $\hat J_x \rightarrow \hat J_x$ and $\hat J_z \rightarrow -\hat J_z$.
The first two terms $\Omega \hat{J_x}$ and $\chi \hat{J}_z^2$, which have even parity symmetry, are invariant under the transformation of exchanging $\left|\uparrow\right\rangle$ and $\left|\downarrow\right\rangle$.
The last term  $\delta \hat{J_z}$, which has odd parity symmetry, changes according to $\delta \hat J_z \rightarrow -\delta \hat J_z$ when $\left|\uparrow\right\rangle \rightarrow \left|\downarrow\right\rangle$.
For an initial state $|\psi\rangle_0=\sum_{m=-J}^J C_m(0) |J,m\rangle$ with $C_m(0)=C_{-m}(0)$ (e.g. $|\psi\rangle_0=|\pi/2,0\rangle_{SCS}$), when $\delta=0$, the evolved state will always possess even parity symmetry (that is $C_m(t)=C_{-m}(t)$)~\cite{SM}.
Thus, the half population difference $\langle \hat J_z(t)\rangle=\sum_{m=-J}^J m |C_m(t)|^2=0$ at the on-resonance point $\delta=0$.
However, if $\delta\neq0$, $\delta \hat J_z$ and $-\delta \hat J_z$ rotate along opposite directions, which results in the antisymmetry property $\langle \hat J_z(\delta, t)\rangle=-\langle \hat J_z(-\delta, t)\rangle$.

Theoretically, one can write the time-evolution in the interaction picture
\begin{equation}\label{Evo_I}
	i \frac{d}{dt}|\psi(t)\rangle^{I}= \hat{H}_I |\psi(t)\rangle^{I},
\end{equation}
where $\hat H_I=e^{i\hat H_0 t}\Omega \hat J_x e^{-i \hat H_0 t}=\frac{\Omega}{2} \left(\hat{J}_{+}^{I} + \hat{J}_{-}^{I}\right)$ with $H_0=\chi\hat{J}_z^2+\delta \hat J_z$ and $\hat{J}_{\pm}^{I}=\hat{J}_{\pm} e^{i\chi t} e^{\pm i (\delta t +2\chi t \hat{J}_z)}$.
$|\psi(t)\rangle^{I}=e^{i \hat H_0 t} |\psi(t)\rangle=\sum_{m=-N/2}^{N/2} C_{m}^{I}(t)|J, m\rangle$ with $|\psi(t)\rangle=\sum_{m=-N/2}^{N/2} C_{m}(t)|J, m\rangle$.
Substituting $|\psi(t)\rangle^{I}$ into Eq.~\eqref{Evo_I}, one can find ${C}_{m}^{I}(\delta, t)={C}_{-m}^{I}(-\delta, t)$ and the population difference  $\langle \hat J_z(\pm \delta, t)\rangle = \pm \sum_{m=-J}^{J} m |C_m^I(\delta, t)|^2$ if $C_m(0)=C_{-m}(0)$
 (see the detailed proof~\cite{SM}).
Thus spectrum is antisymmetric versus the detuning.
%

\begin{figure}[htb]
\centering
\includegraphics[width=\columnwidth]{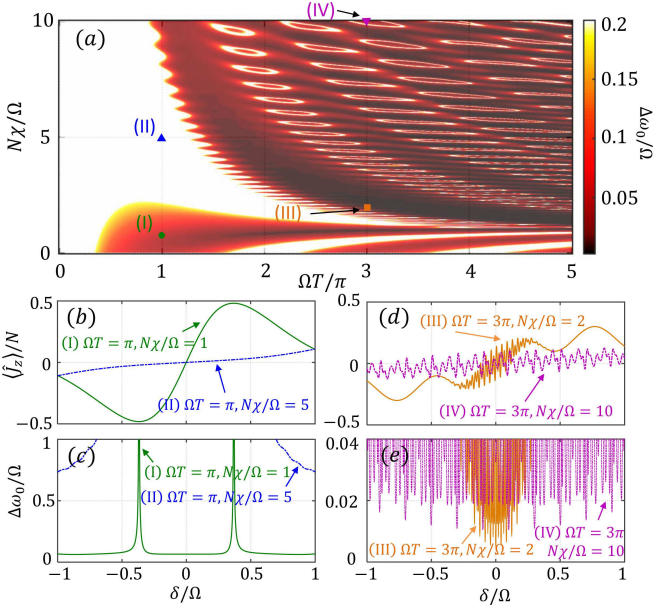}\caption{(Color online) (a) The measurement precision $\Delta\omega_0$ at the on-resonance point versus the evolution time $T$ and the atom-atom interaction $\chi$ for a fixed Rabi frequency $\Omega$. Here, $\Omega=1$. (b) The scaled population difference $\langle\hat{J_z}(T)\rangle/N$ and (c) the measurement precision $\Delta \omega_0$ versus the detuning $\delta$ for the points I (green circle, $T=\pi, \chi=1/N$), II (blue triangle, $T=\pi, \chi=5/N$). (d) The scaled population difference $\langle\hat{J_z}(T)\rangle/N$ and (e) the measurement precision $\Delta \omega_0$ versus the detuning $\delta$ for the points III (orange square, $T=3\pi, \chi=2/N$), IV (purple reversed triangle, $T=3\pi, \chi=10/N$). Here, the total atom number $N=100$ and the initial state $|\psi\rangle_0=|\pi/2,0\rangle_{SCS}$.}
\label{Fig3}
\end{figure}

\begin{figure}[htb]
\centering
\includegraphics[width=\columnwidth]{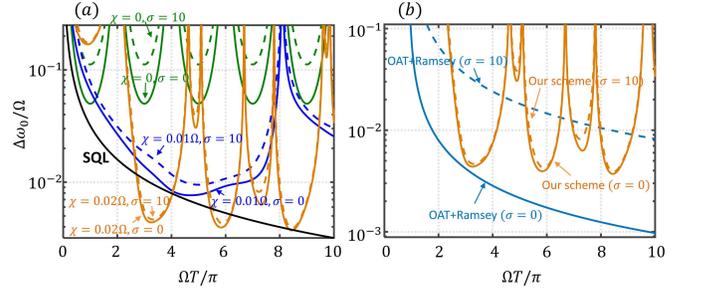}\caption{(Color online) The robustness against detection noises. Here, $\sigma$ characterizes the detection noise strength. (a) The measurement precision $\Delta\omega_0$ versus the evolution time $T$ for a fixed Rabi frequency $\Omega=1$ without and with detection noises. The solid and dashed lines correspond to our scheme with $\sigma=0$ and $\sigma=\sqrt{N}$, respectively. The black solid line denotes the  standard quantum limit. (b) Comparison between our scheme and the Ramsey scheme with spin squeezing. For the Ramsey scheme, we consider the optimal squeezing with $T_p=3^{1/6} N^{-2/3}/\chi$ and the interrogation time $T_R=T-T_p$. Here, for both our scheme and the Ramsey scheme, $\Omega=1$ and $\chi=2/N$ with the total atom number $N=100$.}
\label{Fig4}
\end{figure}

\textit{Interaction-enhanced sensing via antisymmetric Rabi spectroscopy.--}
We choose the non-entangled SCS $|\psi\rangle_0=|\pi/2,0\rangle_{SCS}$ as the input state.
First, for sensors based on Rabi spectroscopy, such as optical lattice clocks~\cite{Takamoto2005, Ludlow2015, Norcia2019, PhysRevLett.128.073603, Liu_2017, Luo_2020}, our antisymmetric Rabi spectroscopy with slight atom-atom interaction can enhance the measurement precision.
In Fig.~\ref{Fig2}~(b), the measurement precision at the on-resonance point with atom-atom interaction (point II, $\chi=0.004\pi$ with $N=100$) is better than the one without atom-atom interaction (point I, $\chi=0$) in the same condition of $\Omega T=\pi$.
The scaled half population difference with atom-atom interaction exhibits sharper slope at the antisymmetric point, resulting in the better measurement precision, see Fig.~\ref{Fig2}~(c) and (d).

Second, for a fixed evolution time $T=1$, as Rabi frequency $\Omega$ and atom-atom interaction $\chi$ increase, the measurement precision at the on-resonance point may be further enhanced.
For example with $\Omega T=3\pi$ and $\chi=0.06\pi$ [the point III in Fig.~\ref{Fig2}~(a)], the measurement precision is much higher than the ones of points I and II in Fig.~\ref{Fig2}~(b).
The signal $\langle \hat J_z (T)\rangle$ exhibits a series of oscillations around the on-resonance point resulting in high resolution and measurement precision, as shown in Fig.~\ref{Fig2}~(c) and (d).
Further, we fix Rabi frequency $\Omega=1$ and show how the measurement precision depends on atom-atom interaction $\chi$ and evolution time $T$, see Fig.~\ref{Fig3}~(a).
For a fixed $\chi$, the measurement precision oscillates with evolution time and there exists an optimal evolution time $T$.
The dependences of $\Delta\omega_0$ on $T$ with different $\chi$ are also shown in Fig.~\ref{Fig4}~(a).

Despite atom-atom interaction can enhance the measurement precision, however $\chi$ cannot be too large.
When $\chi$ is large, $\chi \hat J_z^2$ dominates the Hamiltonian~\eqref{Ham_rabi2} the final population difference will become nearly independent on detuning and the measurement precision becomes bad  dramatically.
For example with $\Omega T=\frac{\pi}{2}$, the strong $\chi$ flattens $\langle \hat J_z (T)\rangle$ versus $\delta$ and the measurement precision decreases a lot, see Fig.~\ref{Fig3}~(b) and (c).
While for $\Omega T=3\pi$, although there have small oscillations, $\langle \hat J_z (T) \rangle$ with strong $\chi$ versus $\delta$ also becomes flat and the measurement precision becomes low, see Fig.~\ref{Fig3}~(d) and (e).
%

\textit{Robustness against imperfect state preparation and detection noises. --}
To implement the antisymmetric Rabi spectroscopy, one needs to prepare the input SCS $|\psi\rangle_0=|\pi/2,0\rangle_{SCS}=e^{i\frac{\pi}{2}\hat J_y}|\psi\rangle_0=|\pi,0\rangle_{SCS}$, which can be easily prepared by a short $\pi/2$ pulse along $y$ axis onto the state of all atoms in $\left|\downarrow\right\rangle$.
We find that even under imperfect $e^{i\frac{1+\epsilon}{2}\pi\hat J_y} $ pulse with $|\epsilon|=0.1$,  it can still keep work~\cite{SM}.
Taking into account the atom-atom interaction, as long as the Rabi frequency of $\pi/2$ pulse $\Omega_{pul} \sim \frac{N^2 \chi}{2}$, the antisymmetric signal can be observed (more details in~\cite{SM}).

In practice, the detection may become imperfect due to the presence of unavoidable detection noises.
Considering the Gaussian detection noise $\sigma^2$, the variance of the population difference becomes $\Delta^2 \tilde{J}_z = \Delta^2 \hat{J}_z + \sigma^2$ and  according to Eq.~\eqref{precision} the measurement precision is $\Delta \omega_0 = {\Delta \tilde{J}_z}/{|\partial \langle \hat J_z(T)\rangle/\partial \omega_0|}$.
In Fig.~\ref{Fig4}~(a), we show the robustness against detection noise within our scheme for $N=100$.
Even under the shot noise $\sigma=\sqrt{N}=10$, the signals with $\chi$ change little compared to the ones of $\sigma=0$, which indicates the strong robustness against detection noise.
This is because the atom-atom interaction in our scheme simultaneously magnifies the quantum fluctuation and the slope of the signal.
When the quantum fluctuation submerges the classical noise, our scheme may achieve excellent robustness against noises~\cite{Nolan2017,PhysRevA.98.030303,Huang2018-2}.

Our antisymmetric Rabi spectrum can exhibit better robustness against detection noises than the Ramsey scheme with spin squeezing~\cite{Wineland1992, Wineland1994, Louchet-Chauvet2010}.
To compare with our scheme, we consider the input state of the Ramsey spectroscopy is an optimal spin squeezed state generated by one-axis twisting (OAT) during $T_p=3^{1/6} N^{-2/3}/\chi$~\cite{Kitagawa1993}.
To make optimal measurement, a rotation $e^{i\alpha\hat{J}_x}$ with $\alpha$ the optimal rotation angle has to be applied~\cite{Ma2011,Gross2012-1}.
Then the state undergoes phase accumulation within a duration of $T_R$.
To compare under the same temporal resource, we set $T_R=T-T_p$.
Finally, a $\pi/2$ pulse along $x$ axis is applied and the population difference measurement is performed.
Unlike Ramsey schemes whose measurement precision increases with interrogation time, the measurement precision in our scheme oscillates with the evolution time and we can find the optimal $T$ numerically.
However, with detection noise ($\sigma=\sqrt{N}$) the measurement precision in Ramsey scheme decreases dramatically while ours remains almost the same.
Our scheme does not need to prepare specific entangled state and to perform nonlinear detection~\cite{Davis2016,Nolan2017,PhysRevA.98.030303,Huang2018-1}, which is much simpler compared to the protocols of Ramsey interferometry with other entangled states~\cite{Hosten2016-2, Li2021-2}.
%

%
%
%
%
%
%
%
%

\textit{Summary and discussions.--}
We have presented a novel protocol for achieving robust interaction-enhanced sensing via antisymmetric Rabi spectroscopy.
%
%
The antisymmetric Rabi spectroscopy have many advantages compared with the conventional ones:
(i) It has no collision shift in the presence of atom-atom interaction;
(ii) Its measurement precision can be improved by using atom-atom interaction;
and (iii) It is robust against detection noise.
Our scheme does not require large modification to the existing experimental setups.
It can be easily realized with state-of-the-art techniques in Bose-condensed atomic systems~\cite{Gross2010, Riedel2010, Ockeloen2013, Strobel2014}, cavity systems with light-mediated interactions~\cite{Davis2016,Colombo2022,Pedrozo-Penafiel2020,PRXQuantum.3.020308,Greve2022}, trapped ions~\cite{Gilmore2021} and etc (The detailed results based on realistic experiments are shown in~\cite{SM}).
Our work revolutionizes the Rabi spectroscopy with higher resolution, measurement precision and robustness, which offers a new way for high-precision frequency measurement.

\begin{acknowledgments}{We thank Dr. Chengyin Han for helpful discussions. This work is supported by the National Key Research and Development Program of China (Grant No. 2022YFA1404104), the National Natural Science Foundation of China (12025509, 11874434), the Key-Area Research and Development Program of GuangDong Province (2019B030330001). J. H. is partially supported by the Guangzhou Science and Technology Projects (202002030459).}	
\end{acknowledgments}

\newpage
\appendix
\setcounter{equation}{0}
\setcounter{figure}{0}
\renewcommand{\theequation}{S\arabic{equation}}
\renewcommand{\thefigure}{S\arabic{figure}}

\section{I. Derivation of the half population difference Eq.~(4)}
In order to analytically calculate $\langle \hat J_z(T)\rangle=\langle \psi(T)|\hat{J_z}|\psi(T)\rangle = _0\!\!\langle \psi| e^{i \hat H_R T} \hat{J_z} e^{-i \hat H_R T}|\psi\rangle_0$, we can make use of the formula $e^{\hat A}\hat B e^{-\hat A}=\sum_{n=0}^{\infty}\frac{1}{n!}\left[\hat A^{(n)},\hat B\right]$.
For Hamiltonian~(1), we have
\begin{equation}\label{SM1}
    \left[\hat{H}_R^{(0)}, \hat{J}_z \right]=\hat{J}_z,
\end{equation}
\begin{equation}\label{SM2}
    \left[\hat{H}_R^{(1)}, \hat{J}_z \right]=\left[\hat{H}_R,\hat{J}_z\right]=\Omega T \hat{J}_z,
\end{equation}
\begin{equation}\label{SM3}
    \left[\hat{H}_R^{(2)}, \hat{J}_z \right]=\left[\hat{H}_R,\Omega t \hat{J}_z\right]=-(\Omega t)^2 \hat{J}_z + (\Omega T)(\delta T) \hat{J}_x.
\end{equation}
When proceeding to higher order, we find that $\left[\hat{H}_R^{(n)}, \hat{J}_z \right]$ will only contain the linear operators of $\hat{J}_x$, $\hat{J}_y$, and $\hat{J}_z$.
Thus, we assume $\left[\hat{H}_R^{(n)}, \hat{J}_z \right]=x_n \hat{J}_x + y_n \hat{J}_y + z_n \hat{J}_z$.
Then, we have
\begin{eqnarray}
    &&\left[\hat{H}_R^{(n+1)}, \hat{J}_z \right]=\left[\left[\hat{H}_R^{(n)}, \hat{J}_z \right], \hat{J}_z\right] \nonumber\\
    &=&-(\delta T)y_n \hat{J}_x + (\Omega T z_n -\delta T x_n) \hat{J}_y - (\Omega T y_n) \hat{J}_z.
\end{eqnarray}
The coefficients satisfy
\begin{eqnarray}
    x_{n+1}&=&(\delta T)y_n, \\
    z_{n+1}&=&-(\Omega T)y_n, \\
    y_{n+1}&=&-(\delta T)x_n+(\Omega T)z_n.
\end{eqnarray}
From Eqs.~\eqref{SM1}-\eqref{SM3}, we have $x_0=0, x_1=0, x_2=(\Omega T)(\delta T)$; $y_0=0, y_1=\Omega t, y_2=0$; $z_0=1, z_1=0, z_2=-(\Omega T)^2$.
Using the recursive relations between these coefficients, we obtain
\begin{eqnarray}
    x_{2n+2}&=&-(\Omega^2+\delta^2)T^2 x_{2n}, x_{2n+1}=0\\
    z_{2n+2}&=&-(\Omega^2+\delta^2)T^2 z_{2n}, z_{2n+1}=0\\
    y_{2n+3}&=&-(\Omega^2+\delta^2)T^2 y_{2n+1}, y_{2n}=0
\end{eqnarray}
with $n=0,1,2,...$. Thus, we can analytically give the expressions for the coefficients as
\begin{eqnarray}
    x_{2n+2}&=&(-1)^n\frac{\Omega\delta}{\Omega^2+\delta^2} \left(\sqrt{\Omega^2+\delta^2} T\right)^{2n+2}, \\
    z_{2n+2}&=&(-1)^n\frac{\Omega^2}{\Omega^2+\delta^2} \left(\sqrt{\Omega^2+\delta^2} T\right)^{2n+2}, \\
    y_{2n+1}&=&(-1)^n\frac{\Omega }{\sqrt{\Omega^2+\delta^2}}\left(\sqrt{\Omega^2+\delta^2} T\right)^{2n+1}.
\end{eqnarray}
Finally, we can analytically obtain the sums for the coefficients,
\begin{eqnarray}\label{XYZ}
    X(\Omega,\delta,T)&=& \sum_{k=1}^{\infty}\frac{x_{2k}}{(2k)!} \nonumber\\
    &=& \frac{\Omega\delta}{\Omega^2+\delta^2}\!\left[1-\cos (\sqrt{\Omega^2+\delta^2} T)\right], \\
    Y(\Omega,\delta,T) &=& \sum_{k=1}^{\infty}\frac{y_{2k-1}}{(2k-1)!} \nonumber\\
    &=& \frac{\Omega \sin (\sqrt{\Omega^2+\delta^2} T)}{\sqrt{\Omega^2+\delta^2}}, \\
    Z(\Omega,\delta,T) &=& \sum_{k=0}^{\infty}\frac{z_{2k}}{(2k)!} \nonumber\\
    &=& \frac{\delta^2+ \Omega^2 \cos (\sqrt{\Omega^2+\delta^2} T)}{\Omega^2+\delta^2}.
\end{eqnarray}
Thus, we get the final result of Eq.~(3) for arbitrary initial state,
\begin{eqnarray}\label{anaJz}
	\langle \hat J_z(T)\rangle=X \langle \hat{J_x}\rangle_0 +Y \langle\hat{J_y}\rangle_0 + Z \langle\hat{J_z}\rangle_0,
\end{eqnarray}
with $X=X(\Omega,\delta,T)$, $Y=Y(\Omega,\delta,T)$, $Z=Z(\Omega,\delta,T)$.

\begin{figure*}[htb]
\centering
\includegraphics[width=2\columnwidth]{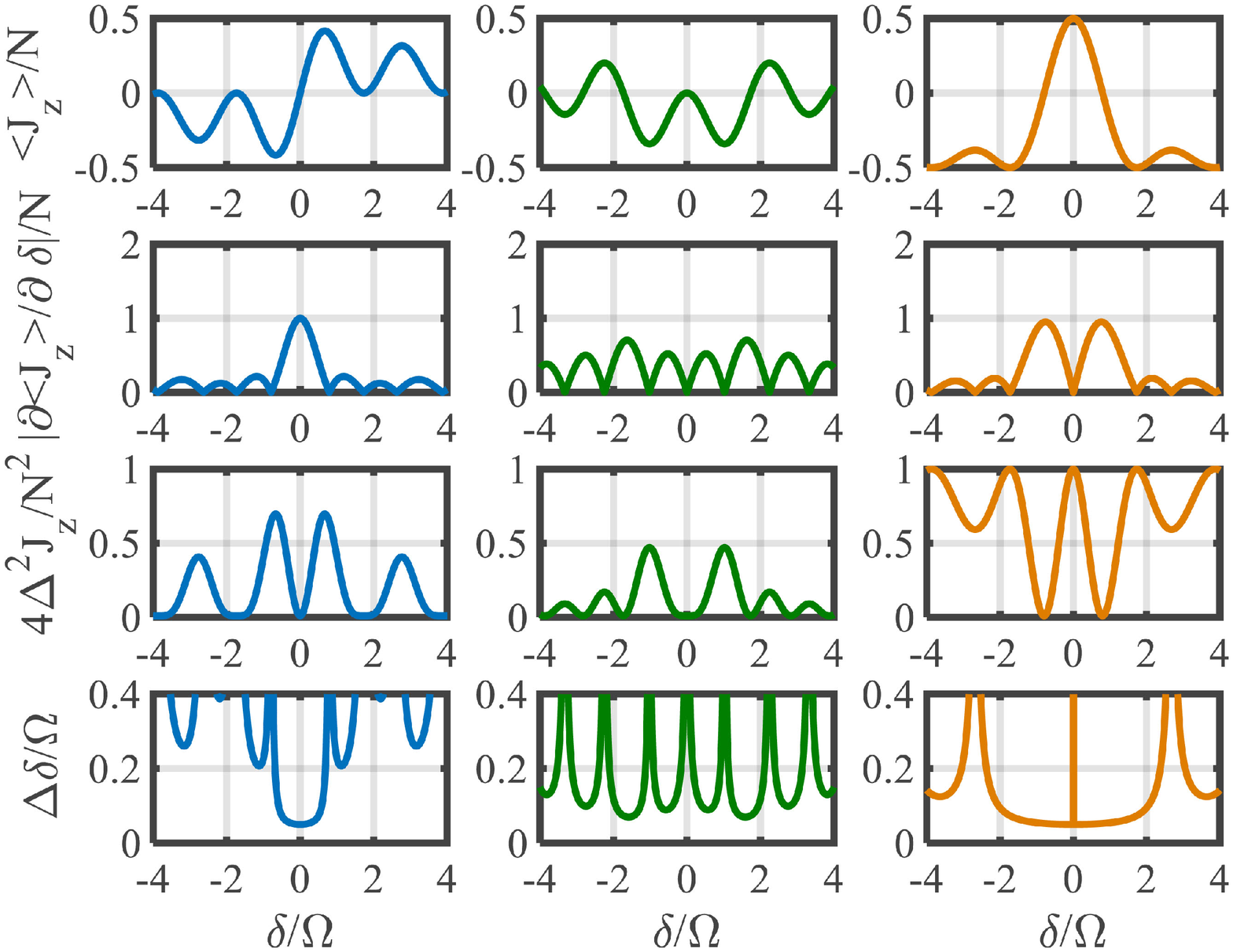}\caption{The first, second and third column represent the results with initial state $|\psi\rangle_0=|\pi/2,0\rangle_{SCS}, |\pi/2,\pi/2\rangle_{SCS}$, and $|\pi,0\rangle_{SCS}$, respectively. The first, second, third and fourth row respectively correspond to the population difference, absolute slope of the population difference, variance of population difference and the measurement precision versus detuning. Here, atom number $N=100$, $\Omega T =\pi$. }
\label{Fig1-SM}
\end{figure*}

\section{II. Measurement precisions for noninteracting systems}

Below we show how to calculate measurement precisions for noninteracting systems (i.e. $\chi=0$).
Given $\delta$ and $\Omega$, we first need to analytically obtain the expressions of $\langle \hat J_z^2(T)\rangle$.
It is easy to find that $\langle \hat J_z^2(T)\rangle=\langle\psi(T)|\hat J_z^2|\psi(T)\rangle=_0\!\!\langle \psi| e^{i \hat H_R T} \hat{J_z^2} e^{-i \hat H_R T}|\psi\rangle_0=_0\!\!\langle \psi| e^{i \hat H_R T} \hat{J_z} e^{-i \hat H_R T} e^{i \hat H_R T} \hat{J_z} e^{-i \hat H_R T}|\psi\rangle_0$.
Thus,
\begin{eqnarray}\label{Jz2}
    \langle \hat J_z^2(T)\rangle &=& _0\langle \psi|\left( X \hat{J}_x + Y \hat{J}_y + Z \hat{J}_z\right)^2 |\psi\rangle_0 \nonumber \\
    &=& XX \langle\hat{J_x^2}\rangle_0 +YY \langle\hat{J_y^2}\rangle_0 +ZZ \langle\hat{J_z^2}\rangle_0 \nonumber \\
    &+& XY \langle\{\hat{J_x},\hat{J_y}\}\rangle_0 +YZ \langle\{\hat{J_y},\hat{J_z}\}\rangle_0 \nonumber \\
    &+& XZ \langle\{\hat{J_x},\hat{J_z}\}\rangle_0,
\end{eqnarray}
where $XX=X^2(\Omega,\delta,T)$, $YY=Y^2(\Omega,\delta,T)$, $ZZ=Z^2(\Omega,\delta,T)$, $XY=X(\Omega,\delta,T)Y(\Omega,\delta,T)$, $YZ=Y(\Omega,\delta,T)Z(\Omega,\delta,T)$, $XZ=X(\Omega,\delta,T)Z(\Omega,\delta,T)$, and $\{\star,\bullet\}=\star\bullet+\bullet\star$ denotes the anti-commutator.
According to Eqs.~\eqref{anaJz} and \eqref{Jz2}, we can explicitly write the expression,
\begin{equation}\label{DeltaJz}
    \Delta \hat J_z = \sqrt{\langle \hat J_z^2(T)\rangle - \langle \hat J_z(T)\rangle^2}.
\end{equation}

Besides, the deviation $\partial \langle \hat J_z(T)\rangle/\partial \delta$ can also be analytically obtained, which reads
\begin{equation}\label{Slope}
    \frac{\partial \langle \hat J_z(T)\rangle}{\partial \delta} = X' \langle \hat{J_x}\rangle_0 +Y' \langle\hat{J_y}\rangle_0 + Z' \langle\hat{J_z}\rangle_0,
\end{equation}
with
\begin{widetext}
\begin{eqnarray}
    X'&=& \frac{\Omega}{\Omega^2+\delta^2}\left[1-\cos (\sqrt{\Omega^2+\delta^2} T)\right] +  \frac{\delta^2 \Omega T}{(\Omega^2+\delta^2)^{3/2}}\left[\sin (\sqrt{\Omega^2+\delta^2} T)\right] \nonumber\\
    &&- \frac{2\delta^2 \Omega }{(\Omega^2+\delta^2)^{2}}\left[1-\cos (\sqrt{\Omega^2+\delta^2} T)\right], \\
    Y'&=& \frac{\delta \Omega T}{\Omega^2+\delta^2} \cos (\sqrt{\Omega^2+\delta^2} T) - \frac{\delta \Omega }{(\Omega^2+\delta^2)^{3/2}} \sin (\sqrt{\Omega^2+\delta^2} T), \\
    Z'&=& \frac{2\delta}{\Omega^2+\delta^2}+\frac{2\delta^3}{(\Omega^2+\delta^2)^2} - \frac{\delta \Omega^2 T}{(\Omega^2+\delta^2)^{3/2}} \sin (\sqrt{\Omega^2+\delta^2} T) \nonumber\\
    &&- \frac{2\delta \Omega^2 }{(\Omega^2+\delta^2)^{2}} \cos (\sqrt{\Omega^2+\delta^2} T).
\end{eqnarray}
\end{widetext}
Finally, according to Eqs.~\eqref{anaJz} - \eqref{Slope},
\begin{equation}\label{precision}
    \Delta \delta = \frac{\Delta \hat J_z}{|\partial \langle \hat J_z(T)\rangle/\partial \delta|}
\end{equation}
can also be analytically expressed. Since $\delta=\omega_0-\omega$, the standard deviation $\Delta\omega_0=\Delta\delta$ mathematically. For convenience, we calculate the measurement precision $\Delta\omega_0$ by $\Delta\delta$ instead.

\section{III. Analytical analysis on antisymmetric locking signal for noninteracting systems}

Below we analyze the antisymmetric locking signal for noninteracting systems.
From Eqs.~\eqref{XYZ} - \eqref{anaJz}, one can easily find that $X(\Omega,\delta,T)=-X(\Omega,-\delta,T)$, $Y(\Omega,\delta,T)=Y(\Omega,-\delta,T)$ and $Z(\Omega,\delta,T)=Z(\Omega,-\delta,T)$ so that $\langle \hat J_z(\delta,T)\rangle=-\langle \hat J_z(-\delta,T)\rangle$ if $\langle\hat{J_y}\rangle_0=\langle\hat{J_z}\rangle_0=0$.
For example, if the initial state is $|\psi\rangle_0=|\pi/2,0\rangle_{SCS}$, $\langle\hat{J_x}\rangle_0=\frac{N}{2}$ and $\langle\hat{J_z}\rangle_0=\langle\hat{J_y}\rangle_0=0$, the final population difference $\langle \hat J_z(T)\rangle$ is antisymmetric with $\delta=0$.
While for $|\psi\rangle_0=|\pi/2,\pi/2\rangle_{SCS}$ and $|\psi\rangle_0=|\pi,0\rangle_{SCS}$ (both with $\langle\hat{J_x}\rangle_0=0$), the final population difference $\langle \hat J_z(T)\rangle$ is symmetric with $\delta=0$.
The comparisons with these three different initial states are shown in the first row of Fig.~\ref{Fig1-SM}.

From Eq.~\eqref{Slope}, we can plot the absolute slope of population difference $|\partial \langle \hat J_z(T)\rangle/\partial \delta|$ versus detuning, see the second row of Fig.~\ref{Fig1-SM}.
There is a peak at $\delta=0$ for $|\psi\rangle_0=|\pi/2,0\rangle_{SCS}$, while the slope equals 0 for both $|\psi\rangle_0=|\pi/2,\pi/2\rangle_{SCS}$ and $|\psi\rangle_0=|\pi,0\rangle_{SCS}$.

From Eq.~\eqref{DeltaJz}, we can also plot the variance of population difference $\Delta^2 \hat{J}_z$ versus detuning, see the third row of Fig.~\ref{Fig1-SM}.
Finally, according to Eq.~\eqref{precision}, we can analytically obtain the measurement precision $\Delta \delta$ versus detuning, see the last row of Fig.~\ref{Fig1-SM}.
It is obviously shown that, only $|\psi\rangle_0=|\pi/2,0\rangle_{SCS}$ can achieve high-precision measurement at the locking point $\delta=0$, while for $|\psi\rangle_0=|\pi/2,\pi/2\rangle_{SCS}$ and $|\psi\rangle_0=|\pi,0\rangle_{SCS}$, the measurement precisions are diverged.

The slope of the signal at the locking point $\delta=0$ can be analytically written as
\begin{equation}\label{Slope}
    \frac{\partial \langle \hat J_z(T)\rangle}{\partial \delta}|_{\delta=0}= \frac{N}{2} \frac{1-\cos \Omega T}{\Omega}.
\end{equation}
When $\Omega T=\pi$, $\frac{\partial \langle \hat J_z(T)\rangle}{\partial \delta}|_{\delta=0}=N/\Omega$ attains its maximum, which indicates the high sensitivity for frequency locking.
At the locking point $\delta=0$, $\Delta \delta \propto 1/\sqrt{N}$, which exhibits the SQL scaling.
\section{IV. Analytical analysis on antisymmetric locking signal for interacting systems}

\begin{figure*}[htb]
\centering
\includegraphics[width=2\columnwidth]{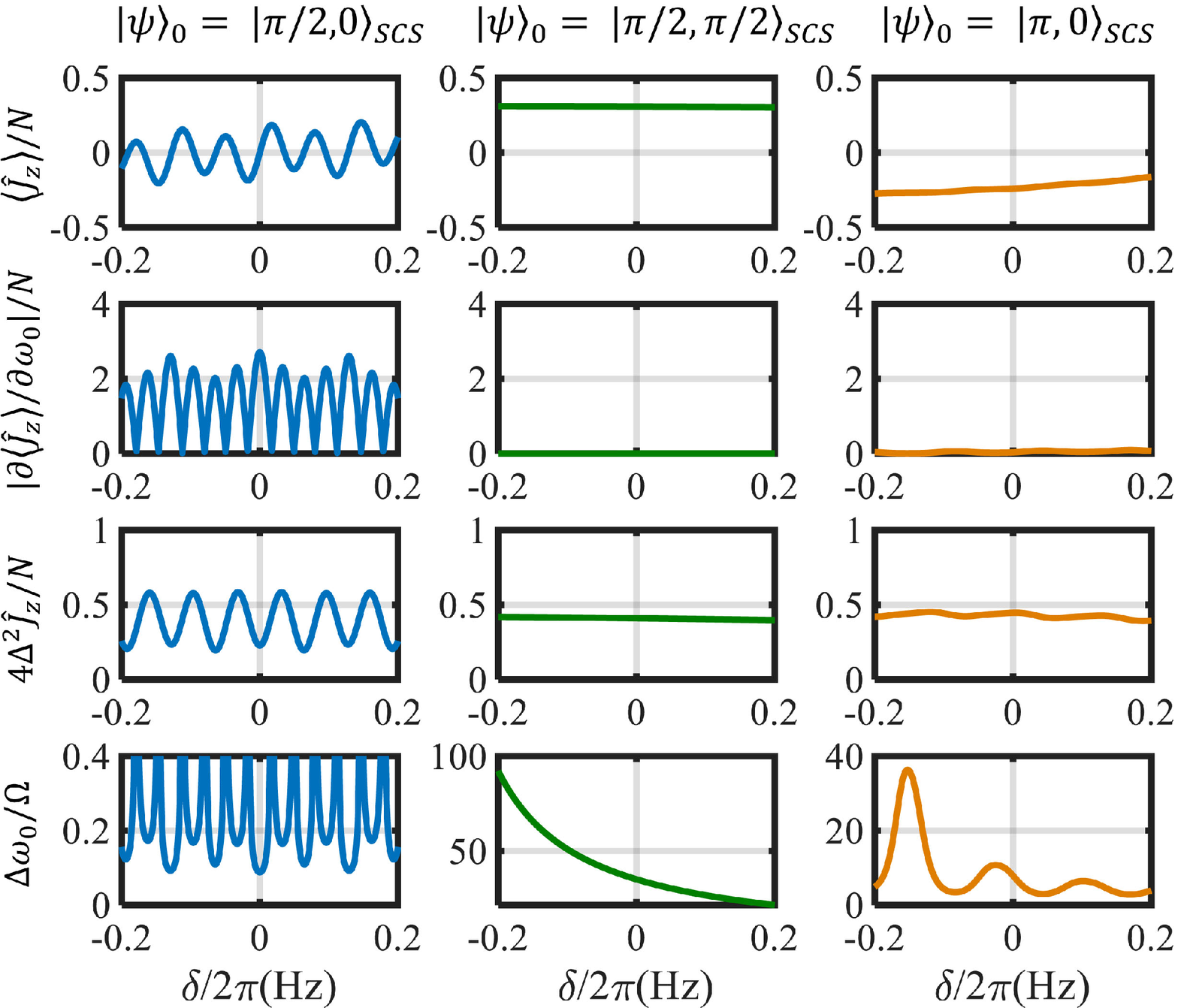}\caption{The first, second and third column represents the results with initial state $|\psi\rangle_0=|\pi/2,0\rangle_{SCS}, |\pi/2,\pi/2\rangle_{SCS}$, and $|\pi,0\rangle_{SCS}$, respectively. The first, second, third and fourth row respectively correspond to the population difference, absolute slope of the population difference, variance of population difference and the measurement precision versus detuning. Here, atom number $N=100$, $\chi=2\pi\times0.063$ Hz, $\Omega=2\pi\times3.15$ Hz, and $T=0.198/\chi=0.5$ s.}
\label{Fig2-SM}
\end{figure*}

In this section, we illustrate the mathematical analysis on the antisymmetric locking signal $\langle \hat J_z(\delta, T)\rangle$ (versus $\delta$) when atom-atom interaction is taken into account.

In the Schrodinger picture, the final state evolved from the initial state $|\psi\rangle_0$ at time $T$ can be calculated as
\begin{equation}\label{Ham_TTS}
	|\psi(T)\rangle_f=e^{i\hat{H}_{R}' T}=e^{-i(\chi\hat{J}_z^2+\Omega \hat{J_x}+\delta \hat J_z)T}|\psi\rangle_{0}.
\end{equation}
In order to analytically analyze the principle of Rabi spectroscopy in the presence of atom-atom interaction, we work in the interaction picture.
In the reference of $\hat H_0=\chi\hat{J}_z^2+\delta \hat J_z$, we have the Hamiltonian
\begin{eqnarray}\label{Ham_I}
	\hat H_{I}&=&e^{i\hat H_0 t}\Omega \hat J_x e^{-i \hat H_0 t} \nonumber \\
&=& \frac{\Omega}{2} e^{i(\chi\hat{J_z^2}+\delta \hat{J_z})t}\left(\hat{J}_{+} + \hat{J}_{-}\right)e^{-i(\chi\hat{J_z^2}+\delta \hat{J_z})t} \nonumber \\
&=& \frac{\Omega}{2} \left(\hat{J}_{+}^{I} + \hat{J}_{-}^{I}\right),
\end{eqnarray}
where $\hat{J}_{\pm}^{I}=e^{i(\chi\hat{J_z^2}+\delta \hat{J_z})t} \hat{J}_{\pm} e^{-i(\chi\hat{J_z^2}+\delta \hat{J_z})t}$.
Since $\frac{d}{dt} \hat{J}_{\pm}^{I}=i\left[\chi\hat{J_z^2}+\delta \hat{J_z}, \hat{J}_{\pm}^{I} \right]$, $\left[\hat{J}_z, \hat{J}_{\pm} \right]=\pm \hat{J}_{\pm}$, and $\left\{\hat{J}_z, \hat{J}_{\pm} \right\}=\hat{J}_{\pm} \left(2\hat{J}_z \pm 1 \right)$, we have
\begin{eqnarray}\label{Jpm_It}
	\frac{d}{dt} \hat{J}_{\pm}^{I}&=&i\left[\chi\hat{J_z^2}+\delta \hat{J_z}, \hat{J}_{\pm}^{I}\right] \nonumber\\
    &=& i\chi\left[\hat{J_z^2}, \hat{J}_{\pm}^{I}\right] + i\delta \left[\hat{J_z}, \hat{J}_{\pm}^{I}\right] \nonumber\\
    &=& \pm i \chi \left\{\hat{J}_z, \hat{J}_{\pm}^{I} \right\} + i\delta \left[\hat{J_z}, \hat{J}_{\pm}^{I}\right] \nonumber\\
    &=& \pm i \hat{J}_{\pm}^{I} \left[\chi\left(2\hat{J}_z \pm 1\right)+\delta\right].
\end{eqnarray}
Therefore, the rasing and lowering operators in the interaction picture can be given as
\begin{equation}\label{Jpm_I}
	\hat{J}_{\pm}^{I}=\hat{J}_{\pm} e^{i\chi t} e^{\pm i (\delta t +2\chi t \hat{J}_z)}.
\end{equation}
The state in interaction picture can be expanded in terms of Dicke basis, i.e.,
\begin{equation}\label{State_I}
	|\psi(t)\rangle^{I}=\sum_{m=-N/2}^{N/2} C_{m}^{I}(t)|J, m\rangle,
\end{equation}
where $|\psi(t)\rangle^{I}=e^{i \hat H_0 t} |\psi(t)\rangle$ with $|\psi(t)\rangle=\sum_{m=-N/2}^{N/2} C_{m}(t)|J, m\rangle$ the evolved state in Schrodinger picture.
Hence, $C_{m}^{I}(t)=e^{i (\chi m^2 +\delta m)t} C_{m}(t)$ and $C_{m}^{I}(0)=C_{m}(0)$.

The time-evolution in interaction picture obeys
\begin{equation}\label{Evo_I}
	i \frac{d}{dt}|\psi(t)\rangle^{I}= \hat{H}_I |\psi(t)\rangle^{I} = \frac{\Omega}{2} \left(\hat{J}_{+}^{I} + \hat{J}_{-}^{I}\right) |\psi(t)\rangle^{I}.
\end{equation}
Owing to the relations $\hat{J}_{+} |J,m\rangle = \sqrt{(J-m)(J+m+1)}|J,m+1\rangle = \lambda_m^{+} |J,m+1\rangle$ and $\hat{J}_{-} |J,m\rangle = \sqrt{(J+m)(J-m+1)}|J,m-1\rangle = \lambda_m^{-} |J,m-1\rangle$, we can get the equations for the coefficients $C_{m}^{I}(t)$. For a given $C_{m}^{I}(t)$, the equation reads
\begin{widetext}

\begin{eqnarray}\label{Cm_I}
	i \dot{C}_{m}^{I}(t) &=& \frac{\Omega}{2} e^{i\chi t} \left[\lambda_{m-1}^{+} C_{m-1}^I(t) e^{i\delta t +2i \chi t(m-1)} + \lambda_{m+1}^{-} C_{m+1}^I(t) e^{-i\delta t -2i \chi t(m+1)} \right]\nonumber\\
    &=& \frac{\Omega}{2} e^{-i\chi t} \left[\lambda_{m-1}^{+} C_{m-1}^I(t) e^{i\delta t +2i \chi t m} + \lambda_{m+1}^{-} C_{m+1}^I(t) e^{-i\delta t -2i \chi t m} \right].
\end{eqnarray}
Since $\lambda_m^{+}=\lambda_{-m}^{-}$, we also have
\begin{eqnarray}\label{Cm_I2}
	i \dot{C}_{-m}^{I}(t) &=& \frac{\Omega}{2} e^{-i\chi t} \left[\lambda_{-m-1}^{+} C_{-m-1}^I(t) e^{i\delta t -2i \chi t m} + \lambda_{-m+1}^{-} C_{-m+1}^I(t) e^{-i\delta t +2i \chi t m} \right] \nonumber\\
&=& \frac{\Omega}{2} e^{-i\chi t} \left[\lambda_{m-1}^{+} C_{-(m-1)}^I(t) e^{-i\delta t +2i \chi t m} + \lambda_{m+1}^{-} C_{-(m+1)}^I(t) e^{i\delta t -2i \chi t m} \right].
\end{eqnarray}

In the following, we verify the antisymmetry property of $\langle \hat J_z(T)\rangle$ versus $\delta=0$.
For fixed $\Omega$ and $\chi$ at time $t$, ${C}_{m}^{I}(\delta, t)$ is dependent on $\delta$.
According to Eqs.~\eqref{Cm_I} and \eqref{Cm_I2}, we have
\begin{eqnarray}\label{Cm_I3}
	i \dot{C}_{m}^{I}(\delta, t) &=& \frac{\Omega}{2} e^{-i\chi t} \left[\lambda_{m-1}^{+} C_{m-1}^I(t) e^{i\delta t +2i \chi t m} + \lambda_{m+1}^{-} C_{m+1}^I(t) e^{-i\delta t -2i \chi t m} \right], \\
    i \dot{C}_{-m}^{I}(-\delta, t) &=& \frac{\Omega}{2} e^{-i\chi t} \left[\lambda_{m-1}^{+} C_{-(m-1)}^I(t) e^{i\delta t +2i \chi t m} + \lambda_{m+1}^{-} C_{-(m+1)}^I(t) e^{-i\delta t -2i \chi t m} \right].
\end{eqnarray}

\end{widetext}

If the coefficients of the initial state $|\psi(0)\rangle=|\psi\rangle_0=\sum_{m=-N/2}^{N/2} C_{m}(0)|J, m\rangle$ satisfy the condition
\begin{equation}\label{CCC}
    C_m(0)=C_{-m}(0)
\end{equation}
for arbitrary $m$, we can immediately find that $\dot{C}_{m}^{I}(\delta, t)=\dot{C}_{-m}^{I}(-\delta, t)$ and therefore
\begin{equation}\label{Cm_It}
	{C}_{m}^{I}(\delta, t)={C}_{-m}^{I}(-\delta, t).
\end{equation}
At time $T$, the final population difference
\begin{eqnarray}\label{Jz_IT}
	\langle \hat J_z(T)\rangle &=& ^{I}\!\langle \psi (T)|\hat{J}_z^{I}|\psi(T)\rangle^{I} = ^{I}\!\!\langle \psi (T)|\hat{J}_z|\psi(T)\rangle^{I} \nonumber\\
&=& \sum_{m=-J}^{J} m |C_m^I(T)|^2.
\end{eqnarray}
For $\delta$, $\langle \hat J_z(\delta, T)\rangle = \sum_{m=-J}^{J} m |C_m^I(\delta, T)|^2$.
While for $-\delta$, $\langle \hat J_z(-\delta, T)\rangle = \sum_{m=-J}^{J} m |C_m^I(-\delta, T)|^2 = \sum_{m=-J}^{J} -m |C_{-m}^I(-\delta, T)|^2 = - \sum_{m=-J}^{J}m |C_m^I(\delta, T)|^2$.
Thus, we finally prove that
\begin{equation}
    \langle \hat J_z(\delta, T)\rangle =-\langle \hat J_z(-\delta, T)\rangle.
\end{equation}

We choose the initial state as $|\psi\rangle_0=|\pi/2,0\rangle_{SCS}=\sum_{m=-J}^J (\frac{1}{2})^{2J} \sqrt{\frac{(2J)!}{(J+m)!(J-m)!}}|J,m\rangle$ in the main text which satisfies the condition~\eqref{CCC}, thus we always get the antisymmetric locking signal for arbitrary final time $T$.
For comparison, we also show the population difference versus detuning with initial states $|\psi\rangle_0=|\pi/2,\pi/2\rangle_{SCS}$, and $|\pi,0\rangle_{SCS}$.
For these two initial states, $C_m(0)$ does not always equal to $C_{-m}(0)$, thus the population difference is neither antisymmetric nor symmetric with respect to $\delta=0$, see the first row in Fig.~\ref{Fig2-SM}.
The absolute slope of the population difference, variance of population difference and the measurement precision versus detuning are also shown in the last three rows of Fig.~\ref{Fig2-SM}.

\section{V. Realization of $\pi/2$ pulse for preparing the initial state}
\begin{figure*}[htb]
\centering
\includegraphics[width=2\columnwidth]{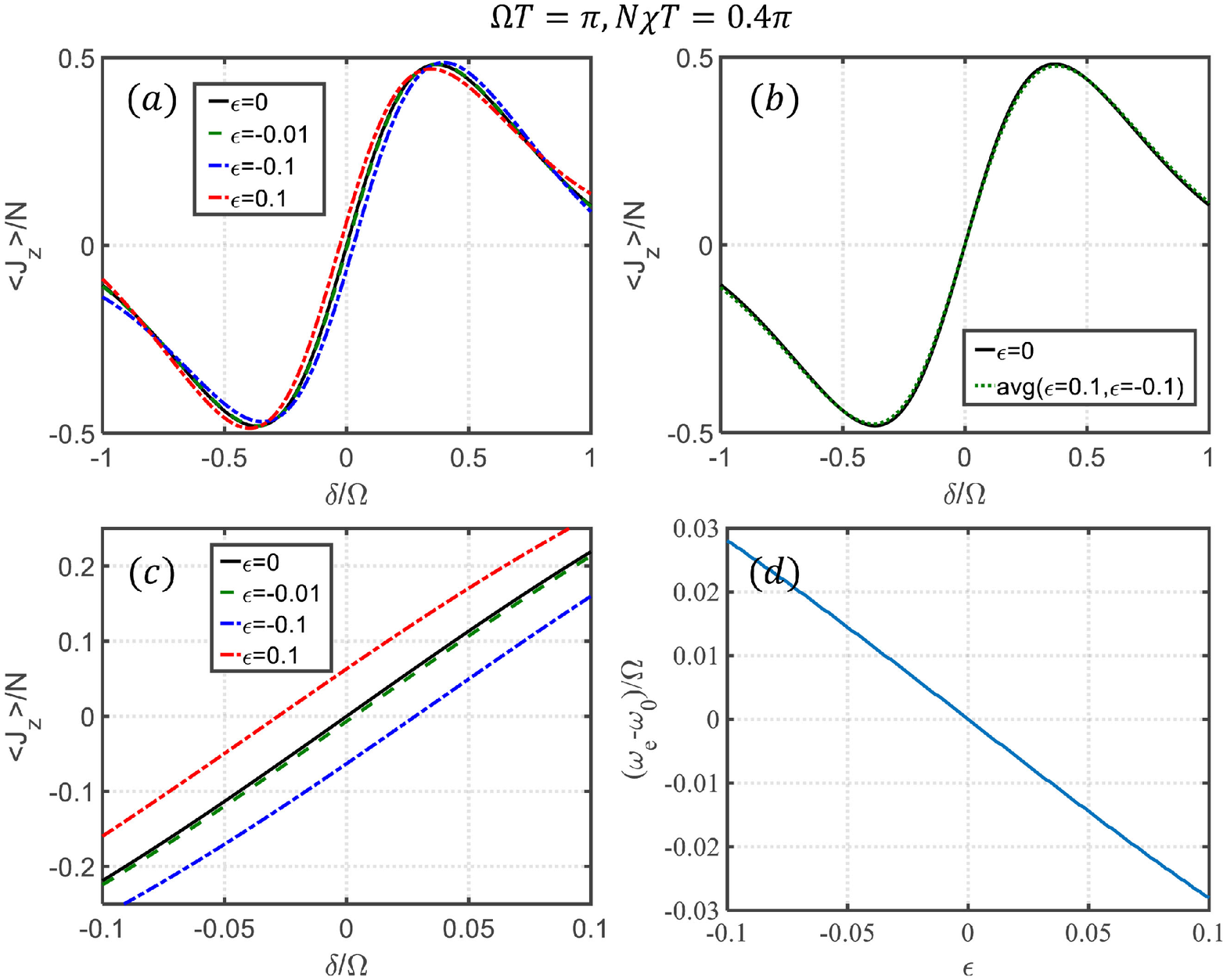}\caption{The locking signal versus detuning with imperfect $\pi/2$ pulse for $\Omega T=\pi$ and $N \chi T=0.4\pi$. The deviation to a perfect $\pi/2$ is characterized by $\epsilon$. (a) The scaled population difference versus detuning under different $\pi/2$ pulse with $\epsilon=0,-0.01,-0.1, 0.1$. (b) The average signal $\frac{1}{2}(\langle\hat{J}_z(\delta,\epsilon)\rangle+\langle\hat{J}_z(\delta,-\epsilon)\rangle)$ with $\epsilon=0.1$, in comparison with the perfect signal. (c) The enlarged area of (a). (d) The shift of antisymmetric point versus deviation $\epsilon$. Here, we choose $N=100$, $\Omega=1$. }
\label{SM-IM1}
\end{figure*}

\begin{figure*}[htb]
\centering
\includegraphics[width=2\columnwidth]{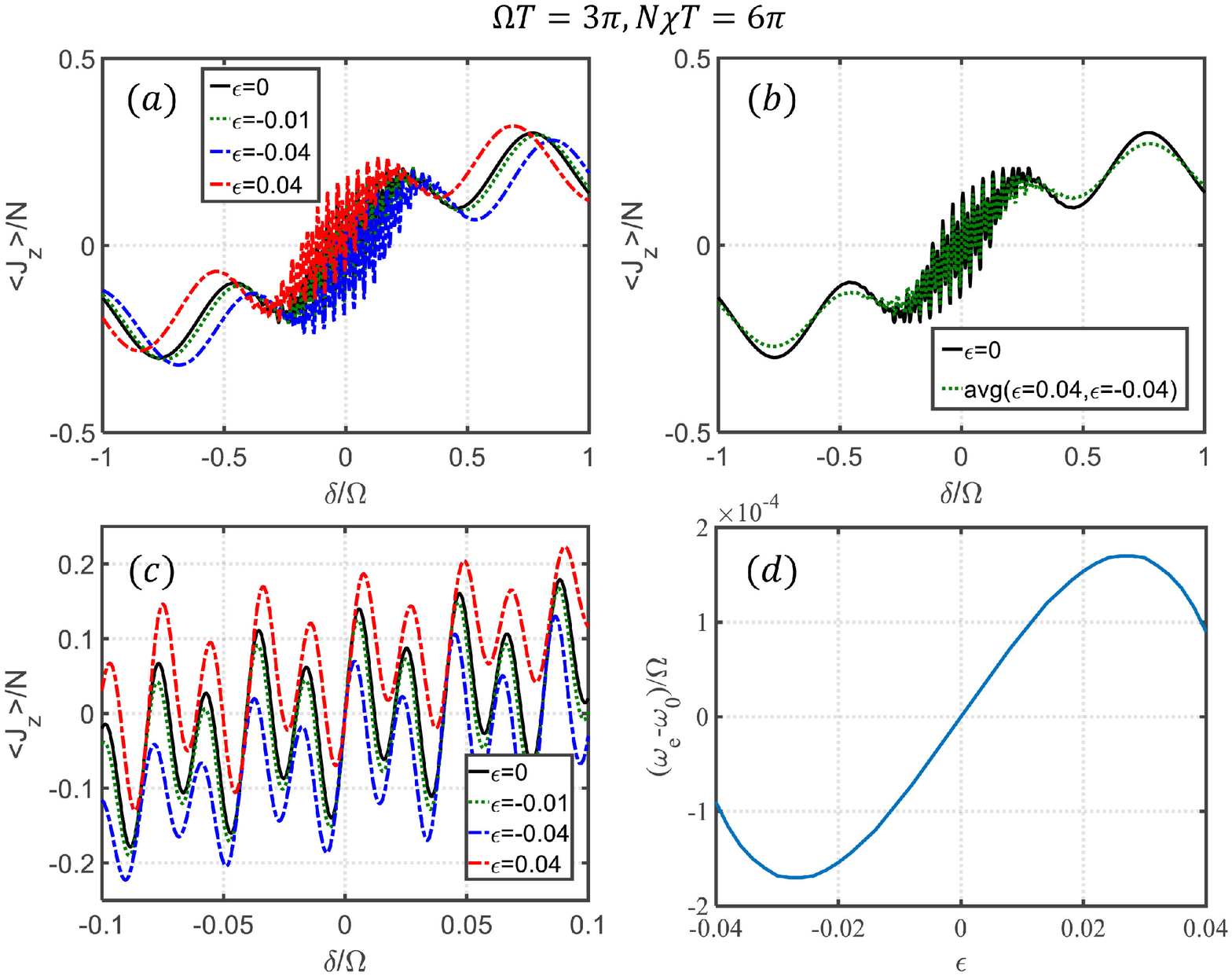}\caption{The locking signal versus detuning with imperfect $\pi/2$ pulse for $\Omega T=3\pi$ and $N \chi T=6\pi$. The deviation to a perfect $\pi/2$ is characterized by $\epsilon$. (a) The scaled population difference versus detuning under different $\pi/2$ pulse with $\epsilon=0,-0.01,-0.04, 0.04$. (b) The average signal $\frac{1}{2}(\langle\hat{J}_z(\delta,\epsilon)\rangle+\langle\hat{J}_z(\delta,-\epsilon)\rangle)$ with $\epsilon=0.04$, in comparison with the perfect signal. (c) The enlarged area of (a). (d) The shift of antisymmetric point versus deviation $\epsilon$. Here, we choose $N=100$, $\Omega=1$.}
\label{SM-IM2}
\end{figure*}

To generate the input state $|\psi\rangle_0=|\pi/2,0\rangle_{SCS}$ for Rabi spectroscopy in our scheme, it is necessary to apply a $\pi/2$ pulse along $y$ axis onto the state of all atoms in $|\downarrow\rangle$.
%
%
%
First, we consider the imperfect $\pi/2$ pulse without atom-atom interaction and detuning, which can be characterized by
\begin{equation}
    e^{i(\frac{1+\epsilon}{2})\pi \hat J_y},
\end{equation}
where $\epsilon$ is the deviation from $\pi/2$. This kind of imperfection mainly comes from the imprecise control of pulse duration.
To better compare with the results in the main text, here we choose $N=100$ for numerical simulation, see Fig.~\ref{SM-IM1} and \ref{SM-IM2}.

When $\epsilon=0$, it is a perfect $\pi/2$ pulse along $y$ axis. Thus, the results for $\epsilon=0$ are consistent with the previous ones.
When $\epsilon\neq0$, the signal of $\langle\hat{J}_z\rangle$ will deviate from the perfect one.
In Fig.~\ref{SM-IM1}~(a) and (c), for $\Omega T=\pi$ and $\chi T=0.4\pi/N$, the signal deviation is not obvious even under $\epsilon=\pm0.1$.
We show the shift of antisymmetric point $\omega_e-\omega_0$ versus $\epsilon$ in Fig.~\ref{SM-IM1}~(d). Here, $\omega_e$ is the antisymmetric point using imperfect pulse with $\epsilon$ and $\omega_0$ is the exact antisymmetric point with perfect pulse.
The shift is only $|\omega_e-\omega_0| \sim 0.03\Omega$ even with $\epsilon=\pm 0.1$.
While for $\Omega T=3\pi$ and $\chi T=6\pi/N$, the signal deviation is still small when $\epsilon=\pm0.04$, see Fig.~\ref{SM-IM2}~(a) and (c).
The shift is only $|\omega_e-\omega_0| \sim 0.0002\Omega$ in the range of $\epsilon\in[-0.04,0.04]$, see Fig.~\ref{SM-IM2}~(d).

However, one can find that the deviations for $\pm \epsilon$ are antisymmetric with respect to $0$, i.e., $\langle\hat{J}_z(\delta,\epsilon)\rangle-\langle\hat{J}_z(\delta,\epsilon=0)\rangle
=-(\langle\hat{J}_z(\delta,-\epsilon)\rangle-\langle\hat{J}_z(\delta,\epsilon=0)\rangle)$, as can be seen in Fig.~\ref{SM-IM1}~(d) and Fig.~\ref{SM-IM2}~(d).
Therefore, the average signal $\frac{1}{2}(\langle\hat{J}_z(\delta,\epsilon)\rangle+\langle\hat{J}_z(\delta,-\epsilon)\rangle)$ becomes antisymmetric with respect to $\delta=0$, see Figs.~\ref{SM-IM1}~(b) and \ref{SM-IM2}~(b).
For nonzero $\chi$, the mean signal differs from the perfect one depending on $\epsilon$ and $\chi$.
However, the antisymmetry will still possess.
Thus in practice, one can scan the pulse duration to obtain the condition for perfect $\pi/2$ pulse ($\epsilon=0$) by analyzing the antisymmetry of the signal.
The influences of the imperfect pulse can be easily eliminated in experiments.

\begin{figure*}[htb]
\centering
\includegraphics[width=2\columnwidth]{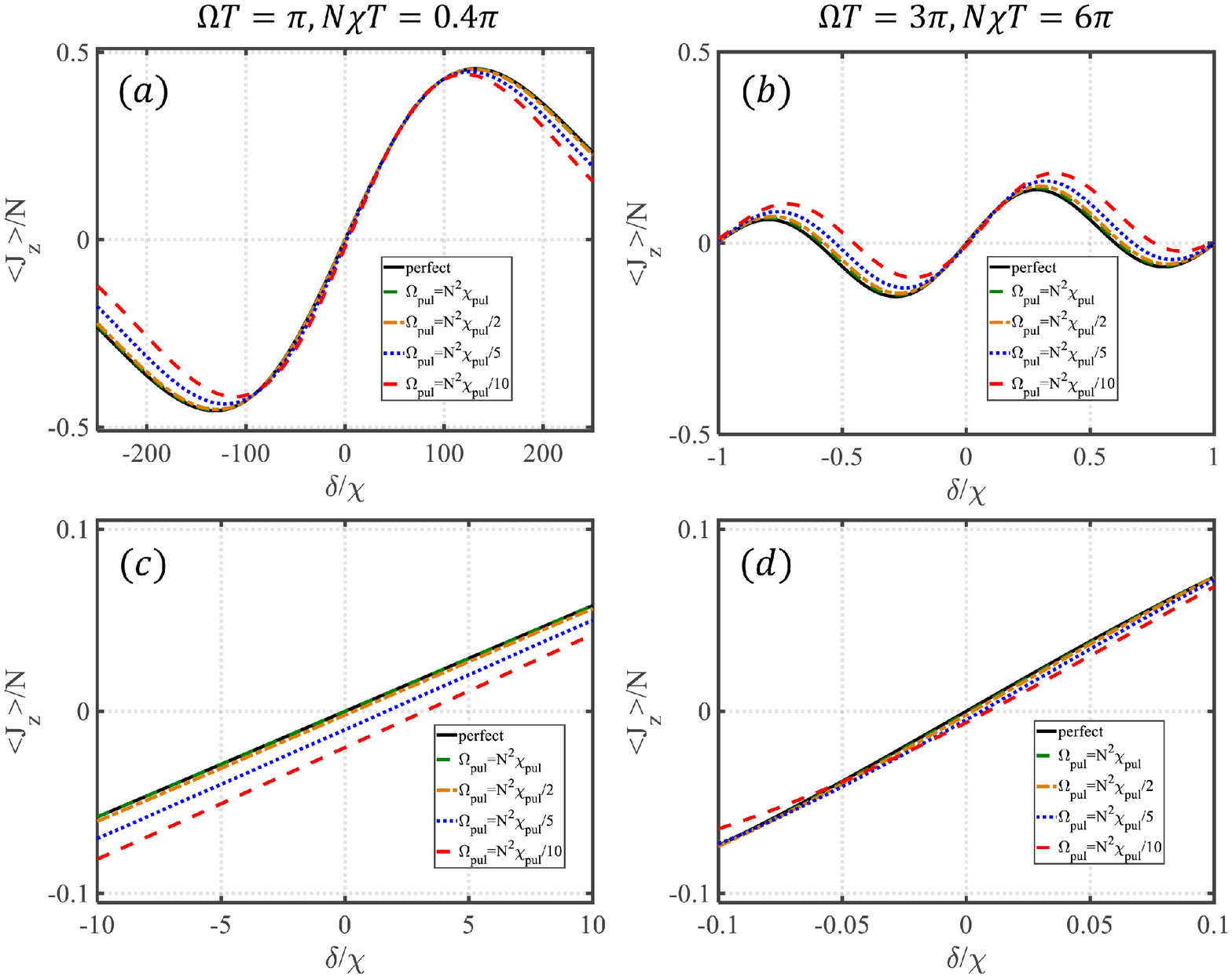}\caption{The final scaled half population difference $\langle\hat J_z\rangle/N$ versus detuning $\delta$ with different imperfect $\pi/2$ pulses. (a) $\Omega T=\pi$, $\chi T=0.4\pi/N$. (b) $\Omega T=3\pi$, $\chi T=6\pi/N$. (c) and (d) are the enlarged region (near on-resonance) of (a) and (b), respectively. Here, we choose $\chi=1$, $\chi_{pul}=\chi$, $N=100$. The black lines are the results with perfect $\pi/2$ pulse. Imperfect $\pi/2$ pulses are chosen with Rabi frequency $\Omega_{pul}=N^2\chi_{pul}, N^2\chi_{pul}/2, N^2\chi_{pul}/5, N^2\chi_{pul}/10$ and pulse duration $t_{pul}=\frac{\pi}{2\Omega_{pul}}$.}
\label{Fig-SM-pulse}
\end{figure*}

In realistic experiments, even the duration of the pulse can be controlled precisely, the atom-atom interaction and the detuning may also affect the $\pi/2$ pulse.
In the following, we take into account the atom-atom interaction and detuning when applying the $\pi/2$ pulse.
The Hamiltonian for generating the $\pi/2$ pulse can be written as
\begin{equation}
    H_{pul}=\chi_{pul} \hat{J}_z^2 + \delta \hat{J}_z - \Omega_{pul}\hat{J}_y,
\end{equation}
where $\chi_{pul}$ and $\Omega_{pul}$ denote the atom-atom interaction strength and the Rabi frequency during the pulse, respectively.
Here, the detuning $\delta$ for $\pi/2$ pulse is the same as the one used in the following Rabi oscillation.
Ideally, if no atom-atom interaction ($\chi_{pul}=0$) and on-resonant with $\delta=0$, a perfect input state can be prepared $|\pi/2,0\rangle_{SCS}=e^{i\frac{\pi}{2}\hat{J}_y}|\pi,0\rangle_{SCS}$ under the condition $\Omega_{pul} t_{pul}=\frac{\pi}{2}$ with $t_{pul}$ the duration of the pulse.

\begin{widetext}
Taking into account $\chi_{pul}$ and $\delta$, the prepared state after the pulse can be expressed as
\begin{equation}
    |\psi\rangle_{pul}=e^{-i H_{pul} t_{pul}} = e^{-i(\chi_{pul} \hat{J}_z^2 + \delta \hat{J}_z - \Omega_{pul}\hat{J}_y) t_{pul}}|\pi,0\rangle_{SCS}.
\end{equation}
Since
\begin{equation}\label{evo_op}
    e^{-i(\chi_{pul} \hat{J}_z^2 + \delta \hat{J}_z - \Omega_{pul}\hat{J}_y) t_{pul}} = e^{-i(\chi_{pul} \hat{J}_z^2 + \delta \hat{J}_z)t_{pul}} e^{i \Omega_{pul}\hat{J}_y) t_{pul}} e^{-\frac{\chi_{pul}\Omega_{pul}t_{pul}^2}{2}[\hat{J}_z^2, \hat{J}_y]-\frac{\delta\Omega_{pul}t_{pul}^2}{2}[\hat{J}_z, \hat{J}_y]} ...,
\end{equation}
If $t_{pul}$ is sufficiently small, Eq.~\eqref{evo_op} can be approximated as
\begin{eqnarray}
    e^{-i(\chi_{pul} \hat{J}_z^2 + \delta \hat{J}_z - \Omega_{pul}\hat{J}_y) t_{pul}} \approx  e^{-i(\chi_{pul} \hat{J}_z^2 + \delta \hat{J}_z)t_{pul}} e^{i \Omega_{pul}\hat{J}_y) t_{pul}}.
\end{eqnarray}
Applying a $\pi/2$ pulse with $\Omega_{pul} t_{pul}=\frac{\pi}{2}$,
\begin{eqnarray}
    e^{-i(\chi_{pul} \hat{J}_z^2 + \delta \hat{J}_z - \Omega_{pul}\hat{J}_y) t_{pul}} \approx  e^{-i\frac{\pi \chi_{pul}}{2\Omega_{pul}} \hat{J}_z^2 -i \frac{\pi \delta}{2\Omega_{pul}} \hat{J}_z} e^{i \frac{\pi}{2} \hat{J}_y}.
\end{eqnarray}
Since $e^{-i\frac{\pi \chi_{pul}}{2\Omega_{pul}} \hat{J}_z^2 -i \frac{\pi \delta}{2\Omega_{pul}} \hat{J}_z}|\frac{N}{2}, m\rangle = e^{-i\frac{\pi \chi_{pul}}{2\Omega_{pul}} m^2 -i \frac{\pi \delta}{2\Omega_{pul}} m}|\frac{N}{2}, m\rangle$, when $\frac{\pi \chi_{pul}}{2\Omega_{pul}}(\frac{N}{2})^2 + \frac{\pi \delta}{2\Omega_{pul}} \frac{N}{2} \rightarrow 0$, $e^{-i\frac{\pi \chi_{pul}}{2\Omega_{pul}} \hat{J}_z^2 -i \frac{\pi \delta}{2\Omega_{pul}} \hat{J}_z}|\frac{N}{2}, m\rangle \approx \hat{I} |\frac{N}{2}, m\rangle = |\frac{N}{2}, m\rangle$ for all $m$.
Therefore, the condition for realizing the $\pi/2$ pulse is
\begin{equation} \label{pulse}
    \Omega_{pul} \gg \frac{\pi N^2 \chi_{pul} + 2\pi \delta N}{8}.
\end{equation}
Thus the prepared state after applying $\pi/2$ pulse becomes
\begin{equation}
    |\psi\rangle_{pul}=e^{-i H_{pul} t_{pul}} \approx e^{i \frac{\pi}{2} \hat{J}_y}|\pi,0\rangle_{SCS}=|\pi/2,0\rangle_{SCS},
\end{equation}
which is the desired SCS input state.
If the $\pi/2$ pulse is perfect, the final signal $\langle \hat J_z(T)\rangle$ is exactly antisymmetric with detuning $\delta$ that can be used for determining the on-resonance frequency.
\end{widetext}

We find numerically that if the Rabi frequency
\begin{equation} \label{pulse2}
    \Omega_{pul} \sim  N^2 \chi_{pul}/2,
\end{equation}
the $\pi/2$ pulse is nearly perfect and the antisymmetric spectrum can still be preserved.
In Fig.~\ref{Fig-SM-pulse}, we show the final signal versus detuning using the input state prepared by $\pi/2$ pulse of different $\Omega_{pul}$.
Here, we assume $\chi_{pul}=\chi=1$ for simulation.
For both cases using $\Omega T=\pi, N\chi T=0.4\pi$ and $\Omega T=3\pi, N\chi T=6\pi$ for Rabi oscillation, the $\pi/2$ pulse satisfying Eq.~\eqref{pulse} can work well.
When $\Omega_{pul}=N^2\chi_{pul}$, the obtained signal is nearly the same with the perfect one.
As $\Omega_{pul}$ decreases, the signal begins to deviate and gradually becomes not antisymmetric with $\delta$.
Despite the antisymmetry breaks down, the zero point $\langle \hat J_z(t)\rangle=0$ can still be used for frequency locking since the deviation is tiny, see the case of $\Omega_{pul}=0.1N^2\chi_{pul}$ in Fig.~\ref{Fig-SM-pulse}~(c) and (d).

\begin{figure*}[htb]
\centering
\includegraphics[width=2\columnwidth]{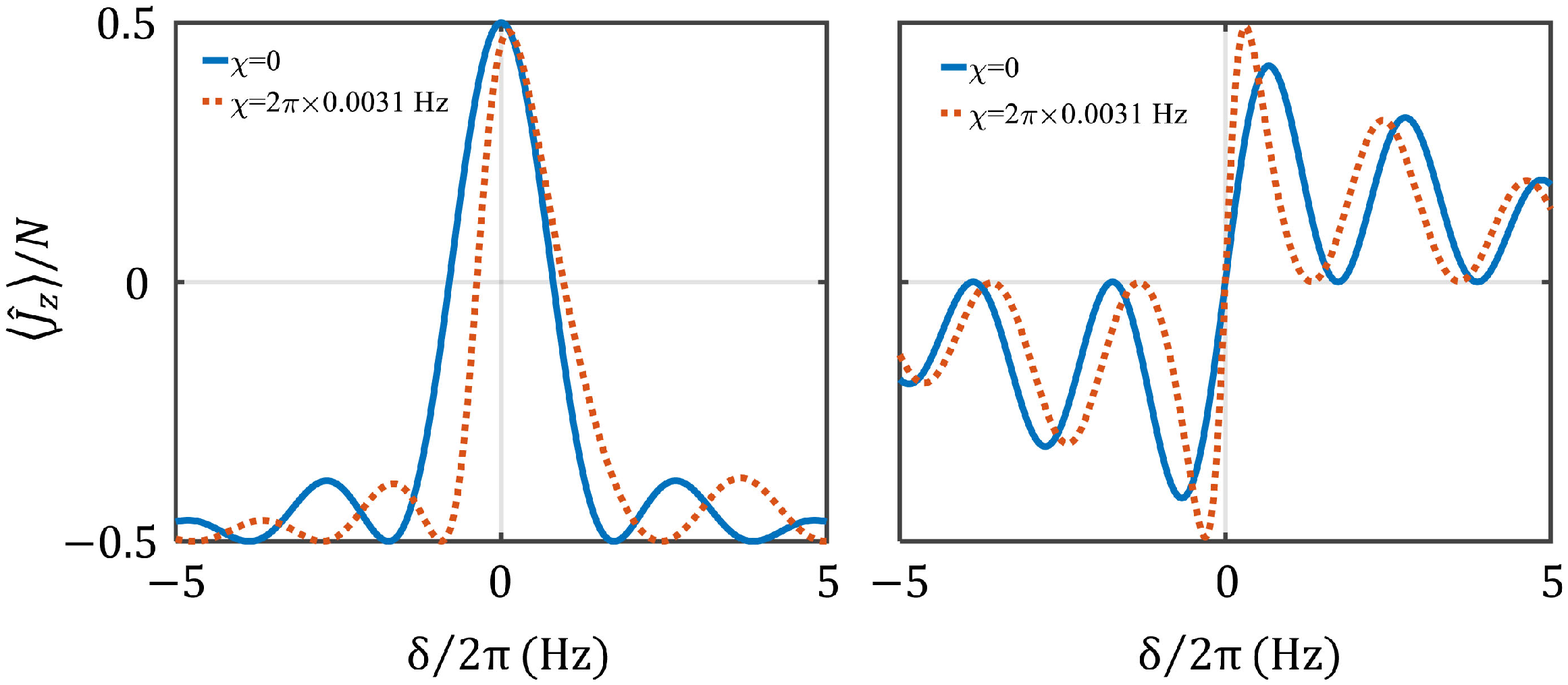}\caption{The conventional Rabi spectroscopy (left) and anti-symmetric Rabi spectroscopy (right) in the presence of atom-atom interaction. Here, the atom number $N=400$, Rabi frequency $\Omega=2\pi\times 1$ Hz and $T=0.5$ s. The blue (solid) and orange (dotted) lines are the results of $\chi=0$ and $\chi=2\pi\times0.0031$ Hz. }
\label{Fig4-SM}
\end{figure*}

\begin{figure*}[htb]
\centering
\includegraphics[width=1.6\columnwidth]{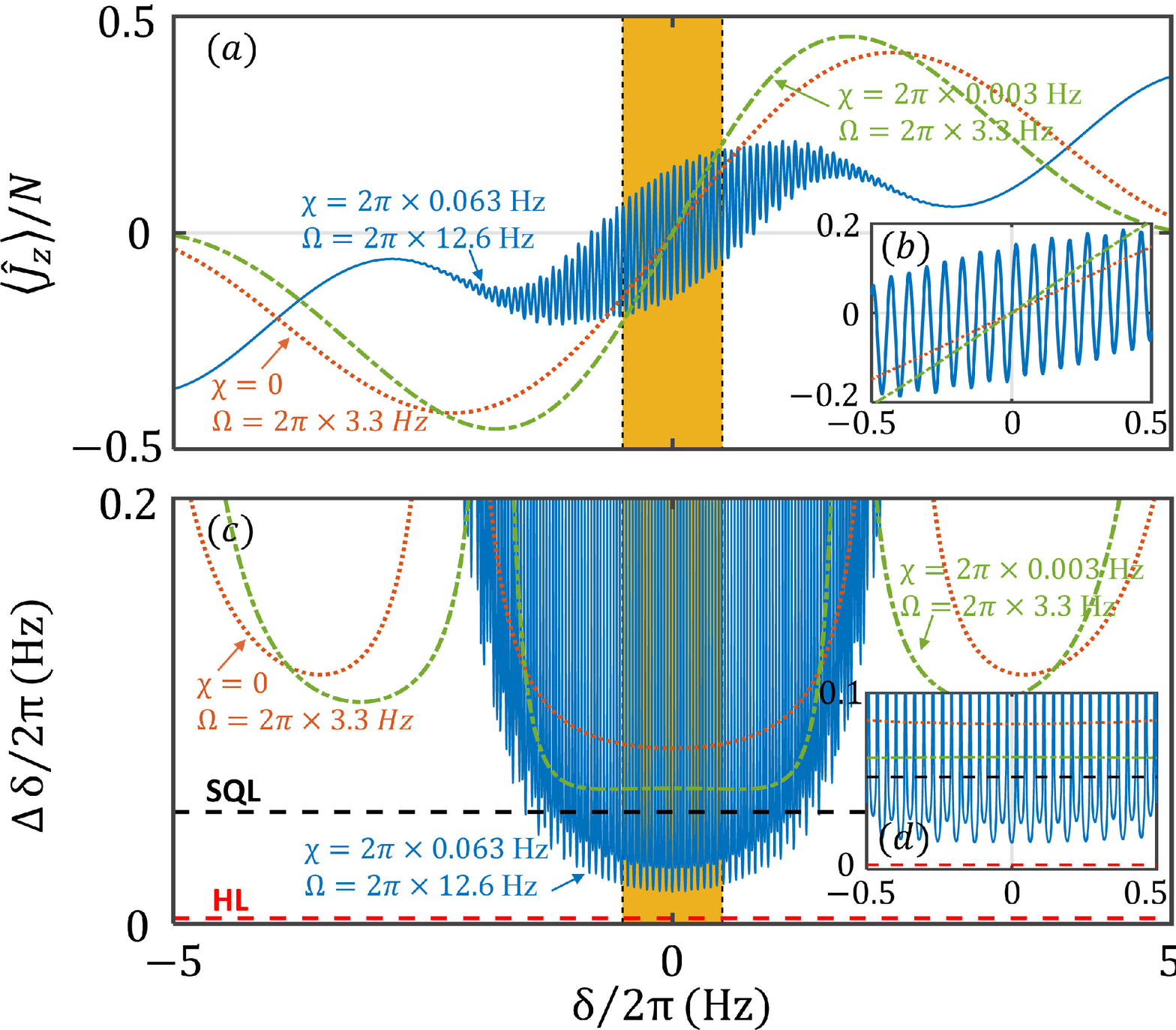}\caption{(a) The scaled population difference $\langle \hat J_z(T)\rangle/N$ versus the detuning $\delta$ and (c) the measurement precision $\Delta\omega_0$ versus $\delta$ for individual atoms ($\chi=0$) with $\Omega=2\pi\times3.3$ Hz, $T=0.1516$ s satisfying the optimal condition $\Omega T=\pi$ and interacting atoms (strong atom-atom interaction $\chi=2\pi\times 0.0063$ Hz with $\Omega=2\pi\times 12.6$ Hz, $T=0.1516$ s ; background atom-atom interaction $\chi=2\pi\times 0.0003$ Hz with $\Omega=2\pi\times 3.3$ Hz, $T=0.1516$ s. (b) and (d) are the insets for the enlarged orange shaded region in (a) and (c), respectively. Here, the total atom number $N=400$ with a perfect initial state $|\psi\rangle_0=|\pi/2,0\rangle_{SCS}$.  }
\label{SM-BEC}
\end{figure*}

\begin{figure*}[htb]
\centering
\includegraphics[width=1.6\columnwidth]{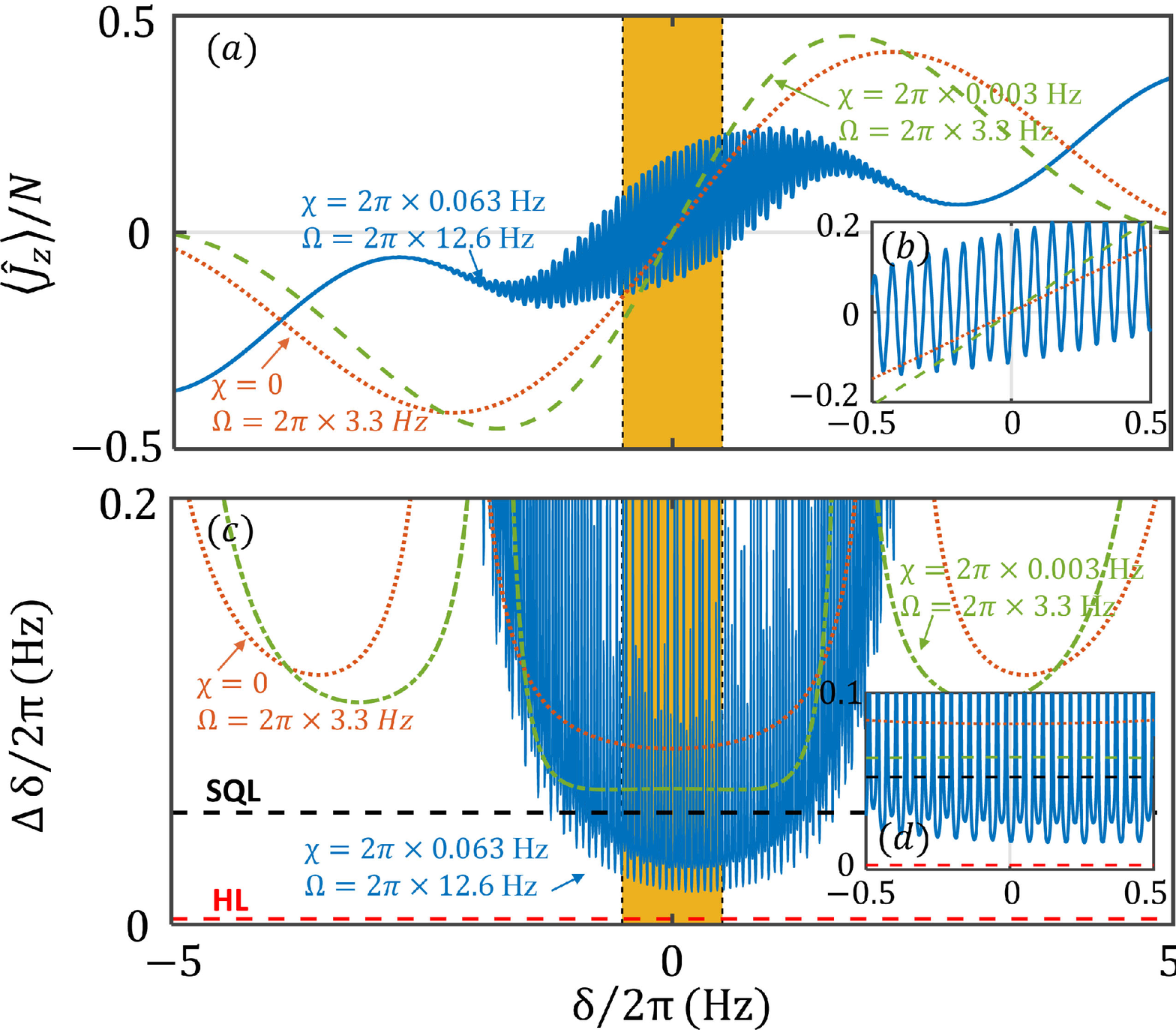}\caption{Simulation results with using imperfect $\pi/2$ pulse for input state preparation. Here, $\Omega_{pul}=2\pi\times 600$ Hz with $t_{pul}=0.416$ ms. The atom-atom interaction is the same during the whole spectroscopy $\chi_{pul}=\chi$. (a) The scaled population difference $\langle \hat J_z(T)\rangle/N$ versus the detuning $\delta$ and (c) the measurement precision $\Delta\omega_0$ versus $\delta$ for individual atoms ($\chi=0$) with $\Omega=2\pi\times3.3$ Hz, $T=0.1516$ s satisfying the optimal condition $\Omega T=\pi$ and interacting atoms (strong atom-atom interaction $\chi=2\pi\times 0.0063$ Hz with $\Omega=2\pi\times 12.6$ Hz, $T=0.1516$ s ; background atom-atom interaction $\chi=2\pi\times 0.0003$ Hz with $\Omega=2\pi\times 3.3$ Hz, $T=0.1516$ s. (b) and (d) are the insets for the enlarged orange shaded region in (a) and (c), respectively. Here, the total atom number $N=400$. }
\label{SM-BEC-im}
\end{figure*}

\section{VI. Experimental feasibility via Bose condensed atoms}
In the following, we take the system of Bose condensed atoms for an example~\cite{Gross2010}.
Another example in collective cavity-QED system is shown in the next section.
We compare the locking signals using our scheme with and without atom-atom interaction.

A Bose-Einstein condensate of $^{87}$Rubidium has been considered as a promising candidate to create coherent spin squeezing based on two hyperfine states~\cite{Gross2010,Gross2012-1,Gross2012-2}.
The $|F,m_F\rangle = |1, 1\rangle$ and $|2, -1\rangle$ states in the lower and upper hyperfine manifold are suitable states for the high-precision measurement experiment.
They fulfill the two major requirements: the tunability of interspecies interactions and their Zeeman energy shifts are to first order common mode with respect to magnetic fields.

The two states are coupled by a two-photon transition comprising of two frequencies in microwave regime and radio-frequency regime, respectively.
The Rabi frequency strength is controlled by the intensity of the electromagnetic radiation.
In the typical Bose condensed atomic system of $^{87}$Rb with atom number $N=400$, for the chosen hyperfine states, the background effective scattering length (in the absence of magnetic field) is given as $a_{aa}+a_{bb}-2a_{ab}\approx 0.5 a_B$ resulting in a nonlinearity $\chi\approx2\pi\times 0.0031$ Hz. The Rabi frequency can be changed from 0 to $2\pi\times600$ Hz.

\subsection{A. Comparison between conventional and antisymmetric Rabi spectroscopy}
For the conventional Rabi spectroscopy, to achieve better measurement precision, the Rabi frequency should be small. Here, we choose $\Omega=2\pi\times 1$ Hz with $T=0.5$ s for simulation.
Considering the background atom-atom interaction, the spectrum has a small shift and the lineshape is no longer symmetric with $\delta=0$, see Fig.~\ref{Fig4-SM}~(a).
This collision shift is harmful for accurately determining the resonance frequency.
In contrast, using the antisymmetric Rabi spectroscopy, the spectrum is always antisymmetric with $\delta=0$, see Fig.~\ref{Fig4-SM}~(b).
The resonance point will not alter when atom-atom interaction is taken into account.
Other results for tunable atom-atom interaction are shown in Fig.~2 and Fig.~3 in the main text, in which the measurement precision can reach beyond SQL with suitably chosen atom-atom interaction $\chi$, Rabi frequency $\Omega$ and evolution time $T$.

\subsection{B. Interaction-enhanced sensing via antisymmetric Rabi spectroscopy}

We choose $\chi=2\pi\times0.063$ Hz, which is a typical atom-atom interaction strength for one-axis twisting with the total atom number $N\approx400$ in experiment~\cite{Gross2010,PhysRevA.86.063623}.
The Rabi frequency is chosen as $\Omega=N\chi/2=2\pi\times12.6$ Hz, which is the optimal condition for twist-and-turn dynamics~\cite{Muessel2015,Mirkhalaf2018,Sorelli2019} whose Hamiltonian is similar to Eq.~(3) in the main text.
For $N=400$, we numerically find the optimal evolution time $T=0.06/\chi\approx0.1516$ s.
In comparison with the one via non-interacting atoms, we choose $T=0.1516$ s and $\Omega=2\pi\times3.3$ Hz which satisfy the optimal condition $\Omega T=\pi$.
Meanwhile, we also consider the case with background atom-atom interaction $\chi=2\pi\times0.003$.

As shown in Fig.~\ref{SM-BEC}~(a) and (b), in the presence of background atom-atom interaction, the measurement precision can be improved compared to the non-interacting case.
While with stronger atom-atom interaction and Rabi frequency ($\chi=2\pi\times0.063$ and $\Omega=2\pi\times12.6$), the locking signal with  oscillates dramatically (especially near $\delta=0$).
Unlike the conventional Rabi spectroscopy, the resolution of the antisymmetric Rabi spectroscopy can be greatly enhanced in the presence of atom-atom interaction.
The slope at the on-resonance point is much sharper which indicates better sensitivity for frequency locking.
In Fig.~\ref{SM-BEC}~(c) and (d), the measurement precisions are shown.
Here, for the same evolution time $T$, the measurement precision with $\chi>0$ is much better than the ones with $\chi=0$.
With optimized $\chi$ and $\Omega$, the measurement precision can surpass the SQL, $(\Delta\omega_0)_{SQL} \propto 1/(\sqrt{N}T)$.

\subsection{C. Influences of imperfect $\pi/2$ pulse for preparing the initial state}
To obtain the desired input state for antisymmetric Rabi spectroscopy, one needs to use a short $\pi/2$ pulse with large Rabi frequency $\Omega_{pul}$.
In this experimental system, the Rabi frequency can be up to $\Omega_{pul}=2\pi\times 600$ Hz.
For small atom-atom interaction $\chi=2\pi\times0.003$ Hz, $\Omega_{pul}=2\pi\times 600$ Hz is much larger than $N^2 \chi$, which satisfies the condition~\eqref{pulse}, see the dashed line in Fig.~\ref{SM-BEC-im}~(a) and (b).
Thus, the signal is exactly antisymmetric with $\delta$ and the measurement precision is the same with the perfect one, see the dashed lines in Fig.~\ref{SM-BEC-im}~(c) and (d).
For strong atom-atom interaction $\chi=2\pi\times0.063$ Hz, despite the Rabi frequency $\Omega_{pul}=2\pi\times 600$ Hz does not meet the strong Rabi frequency condition~\eqref{pulse}, we find that it is sufficient to observe the antisymmetric signal and the shift of the antisymmetric point is small.
As shown in Fig.~\ref{SM-BEC-im}~(a), the blue solid line becomes not exactly antisymmetric with respect to detuning.
Looking closer in the vicinity of the on-resonance point, there appears a shift compared to the perfect one.
However, this shift is extremely small, which merely affect the accuracy of determining the on-resonance frequency.
Besides, the measurement precision hardly change compared to the perfect one.
Thus, the antisymmetric Rabi spectroscopy is experimentally feasible with state-of-the-art techniques via Bose condensed atoms.

\section{VII. Experimental feasibility via collective Cavity-QED system}
\begin{figure*}[htb]
\centering
\includegraphics[width=2\columnwidth]{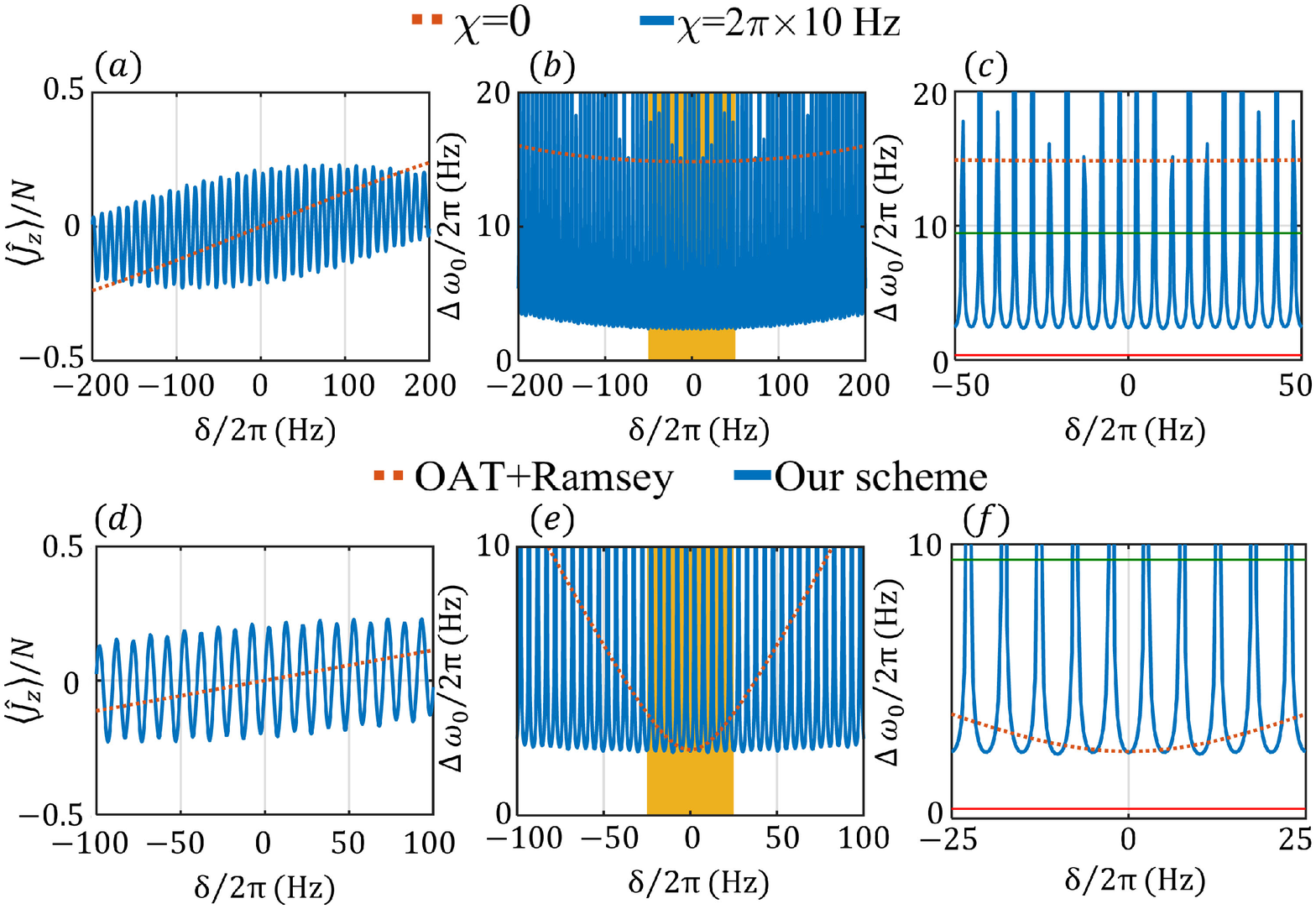}\caption{(a) The final population difference at time $T$ versus $\delta$ and (b) the measurement precision versus $T$ at $\delta=0$ with individual atoms ($\chi=0$) and interacting atoms ($\chi=2\pi \times 10$ Hz). For $\chi=2\pi \times 10$ Hz, we choose $\Omega=2\pi \times 3500$ Hz, $T=0.636$ ms. For $\chi=0$, we choose $\Omega=2\pi \times 785$ Hz, $T=0.636$ ms under the optimal condition $\Omega T=\pi$. (c) is the enlarged orange shaded region in (b).(d) The final population difference at time $T$ versus $\delta$ and (e) the measurement precision versus $T$ at $\delta=0$ using our scheme and Ramsey scheme with spin squeezed state generated via one-axis twisting. For the latter scheme, we consider optimal squeezing with $T_p=3^{1/6} N^{-2/3}/\chi$ and the Ramsey interrogation time $T_R=T-T_p$. For both schemes, $\chi=2\pi \times 10$ Hz, $T=0.636$ ms. (f) is the enlarged orange shaded region in (e). Here, atom number $N=700$.}
\label{Fig5-SM}
\end{figure*}
Our scheme can also be applied to other many-body quantum systems, such as the collective cavity-QED system.
For the atomic ensemble in an optical cavity~\cite{Pedrozo-Penafiel2020,Colombo2022,PRXQuantum.3.020308}, the one-axis twisting can be generated by quantum non-demolition measurements and cavity-mediated spin interaction.
In Ref.~\cite{Greve2022}, the spin squeezed states created by quantum non-demolition measurements and cavity-mediated spin interaction are demonstrated, which provides an ideal platform for achieving the quantum enhanced measurement via antisymmetric Rabi spectroscopy.

In this kind of experimental systems, the interaction strength $\chi$ can be easily tuned via tuning the detuning of the probe laser.
Thus, for the input state preparation stage, one can use a small $\chi_{pul}$ to achieve a perfect $\pi/2$ pulse with $\Omega_{pul} \gg N^2 \chi_{pul}$.
Then, one can use a large $\chi$ for implementing the Rabi oscillations to realize the interaction-enhanced sensing via antisymmetric Rabi spectroscopy.

In the experiment~\cite{Greve2022}, the one-axis twisting strength can be up to $\chi_{OAT}\approx2\pi\times10$ Hz with atom number $N\approx700$.
Thus, in our simulation for implementing Rabi oscillations, we choose $\chi=2\pi\times10$ Hz and $N=700$.
With Rabi frequency $\Omega=N\chi/2=2\pi\times3500$ Hz, the optimal evolution time can be numerically obtained as $T=0.636$ ms.
Under the same evolution time, we choose $\Omega=2\pi\times785$ Hz in the case of $\chi=0$ for comparison.
Besides, we also compare the results of using Ramsey spectroscopy with optimal one-axis twisting spin squeezed state.
The results are shown in Fig.~\ref{Fig5-SM}, which are similar with the ones in Bose condensed atomic system.
According to these results, our protocol can also greatly improves the resolution of the spectrum and the frequency measurement precision in the collective cavity-QED systems.


\begin{thebibliography}{62}%
\makeatletter
\providecommand \@ifxundefined [1]{%
 \@ifx{#1\undefined}
}%
\providecommand \@ifnum [1]{%
 \ifnum #1\expandafter \@firstoftwo
 \else \expandafter \@secondoftwo
 \fi
}%
\providecommand \@ifx [1]{%
 \ifx #1\expandafter \@firstoftwo
 \else \expandafter \@secondoftwo
 \fi
}%
\providecommand \natexlab [1]{#1}%
\providecommand \enquote  [1]{``#1''}%
\providecommand \bibnamefont  [1]{#1}%
\providecommand \bibfnamefont [1]{#1}%
\providecommand \citenamefont [1]{#1}%
\providecommand \href@noop [0]{\@secondoftwo}%
\providecommand \href [0]{\begingroup \@sanitize@url \@href}%
\providecommand \@href[1]{\@@startlink{#1}\@@href}%
\providecommand \@@href[1]{\endgroup#1\@@endlink}%
\providecommand \@sanitize@url [0]{\catcode `\\12\catcode `\$12\catcode
  `\&12\catcode `\#12\catcode `\^12\catcode `\_12\catcode `\%12\relax}%
\providecommand \@@startlink[1]{}%
\providecommand \@@endlink[0]{}%
\providecommand \url  [0]{\begingroup\@sanitize@url \@url }%
\providecommand \@url [1]{\endgroup\@href {#1}{\urlprefix }}%
\providecommand \urlprefix  [0]{URL }%
\providecommand \Eprint [0]{\href }%
\providecommand \doibase [0]{http://dx.doi.org/}%
\providecommand \selectlanguage [0]{\@gobble}%
\providecommand \bibinfo  [0]{\@secondoftwo}%
\providecommand \bibfield  [0]{\@secondoftwo}%
\providecommand \translation [1]{[#1]}%
\providecommand \BibitemOpen [0]{}%
\providecommand \bibitemStop [0]{}%
\providecommand \bibitemNoStop [0]{.\EOS\space}%
\providecommand \EOS [0]{\spacefactor3000\relax}%
\providecommand \BibitemShut  [1]{\csname bibitem#1\endcsname}%
\let\auto@bib@innerbib\@empty
\bibitem [{\citenamefont {Degen}\ \emph {et~al.}(2017)\citenamefont {Degen},
  \citenamefont {Reinhard},\ and\ \citenamefont {Cappellaro}}]{Degen2017}%
  \BibitemOpen
  \bibfield  {author} {\bibinfo {author} {\bibfnamefont {C.~L.}\ \bibnamefont
  {Degen}}, \bibinfo {author} {\bibfnamefont {F.}~\bibnamefont {Reinhard}}, \
  and\ \bibinfo {author} {\bibfnamefont {P.}~\bibnamefont {Cappellaro}},\
  }\bibfield  {title} {\enquote {\bibinfo {title} {Quantum sensing},}\ }\href
  {\doibase 10.1103/RevModPhys.89.035002} {\bibfield  {journal} {\bibinfo
  {journal} {Rev. Mod. Phys.}\ }\textbf {\bibinfo {volume} {89}},\ \bibinfo
  {pages} {035002} (\bibinfo {year} {2017})}\BibitemShut {NoStop}%
\bibitem [{\citenamefont {Kitching}\ \emph {et~al.}(2011)\citenamefont
  {Kitching}, \citenamefont {Knappe},\ and\ \citenamefont {Donley}}]{5778937}%
  \BibitemOpen
  \bibfield  {author} {\bibinfo {author} {\bibfnamefont {John}\ \bibnamefont
  {Kitching}}, \bibinfo {author} {\bibfnamefont {Svenja}\ \bibnamefont
  {Knappe}}, \ and\ \bibinfo {author} {\bibfnamefont {Elizabeth~A.}\
  \bibnamefont {Donley}},\ }\bibfield  {title} {\enquote {\bibinfo {title}
  {Atomic sensors -- a review},}\ }\href {\doibase 10.1109/JSEN.2011.2157679}
  {\bibfield  {journal} {\bibinfo  {journal} {IEEE Sensors Journal}\ }\textbf
  {\bibinfo {volume} {11}},\ \bibinfo {pages} {1749--1758} (\bibinfo {year}
  {2011})}\BibitemShut {NoStop}%
\bibitem [{\citenamefont {Diddams}\ \emph {et~al.}(2004)\citenamefont
  {Diddams}, \citenamefont {Bergquist}, \citenamefont {Jefferts},\ and\
  \citenamefont {Oates}}]{Diddams2004}%
  \BibitemOpen
  \bibfield  {author} {\bibinfo {author} {\bibfnamefont {S.~A.}\ \bibnamefont
  {Diddams}}, \bibinfo {author} {\bibfnamefont {J.~C.}\ \bibnamefont
  {Bergquist}}, \bibinfo {author} {\bibfnamefont {S.~R.}\ \bibnamefont
  {Jefferts}}, \ and\ \bibinfo {author} {\bibfnamefont {C.~W.}\ \bibnamefont
  {Oates}},\ }\bibfield  {title} {\enquote {\bibinfo {title} {{Standards of
  Time and Frequency at the Outset of the 21st Century}},}\ }\href {\doibase
  10.1126/science.1102330} {\bibfield  {journal} {\bibinfo  {journal}
  {Science}\ }\textbf {\bibinfo {volume} {306}},\ \bibinfo {pages} {1318--1324}
  (\bibinfo {year} {2004})}\BibitemShut {NoStop}%
\bibitem [{\citenamefont {Oelker}\ \emph {et~al.}(2019)\citenamefont {Oelker},
  \citenamefont {Hutson}, \citenamefont {Kennedy}, \citenamefont {Sonderhouse},
  \citenamefont {Bothwell}, \citenamefont {Goban}, \citenamefont {Kedar},
  \citenamefont {Sanner}, \citenamefont {Robinson}, \citenamefont {Marti},
  \citenamefont {Matei}, \citenamefont {Legero}, \citenamefont {Giunta},
  \citenamefont {Holzwarth}, \citenamefont {Riehle}, \citenamefont {Sterr},\
  and\ \citenamefont {Ye}}]{Oelker2019}%
  \BibitemOpen
  \bibfield  {author} {\bibinfo {author} {\bibfnamefont {E.}~\bibnamefont
  {Oelker}}, \bibinfo {author} {\bibfnamefont {R.~B.}\ \bibnamefont {Hutson}},
  \bibinfo {author} {\bibfnamefont {C.~J.}\ \bibnamefont {Kennedy}}, \bibinfo
  {author} {\bibfnamefont {L.}~\bibnamefont {Sonderhouse}}, \bibinfo {author}
  {\bibfnamefont {T.}~\bibnamefont {Bothwell}}, \bibinfo {author}
  {\bibfnamefont {A.}~\bibnamefont {Goban}}, \bibinfo {author} {\bibfnamefont
  {D.}~\bibnamefont {Kedar}}, \bibinfo {author} {\bibfnamefont
  {C.}~\bibnamefont {Sanner}}, \bibinfo {author} {\bibfnamefont {J.~M.}\
  \bibnamefont {Robinson}}, \bibinfo {author} {\bibfnamefont {G.~E.}\
  \bibnamefont {Marti}}, \bibinfo {author} {\bibfnamefont {D.~G.}\ \bibnamefont
  {Matei}}, \bibinfo {author} {\bibfnamefont {T.}~\bibnamefont {Legero}},
  \bibinfo {author} {\bibfnamefont {M.}~\bibnamefont {Giunta}}, \bibinfo
  {author} {\bibfnamefont {R.}~\bibnamefont {Holzwarth}}, \bibinfo {author}
  {\bibfnamefont {F.}~\bibnamefont {Riehle}}, \bibinfo {author} {\bibfnamefont
  {U.}~\bibnamefont {Sterr}}, \ and\ \bibinfo {author} {\bibfnamefont
  {J.}~\bibnamefont {Ye}},\ }\bibfield  {title} {\enquote {\bibinfo {title}
  {{Demonstration of 4.8 x 10-17 stability at 1 s for two independent
  optical clocks}},}\ }\href {\doibase 10.1038/s41566-019-0493-4} {\bibfield
  {journal} {\bibinfo  {journal} {Nature Photonics}\ }\textbf {\bibinfo
  {volume} {13}},\ \bibinfo {pages} {714--719} (\bibinfo {year}
  {2019})}\BibitemShut {NoStop}%
\bibitem [{\citenamefont {Orenes}\ \emph {et~al.}(2021)\citenamefont {Orenes},
  \citenamefont {Sewell}, \citenamefont {Lodewyck},\ and\ \citenamefont
  {Mitchell}}]{Orenes2021}%
  \BibitemOpen
  \bibfield  {author} {\bibinfo {author} {\bibfnamefont {Daniel~Benedicto}\
  \bibnamefont {Orenes}}, \bibinfo {author} {\bibfnamefont {Robert~J.}\
  \bibnamefont {Sewell}}, \bibinfo {author} {\bibfnamefont
  {J{\'{e}}r{\^{o}}me}\ \bibnamefont {Lodewyck}}, \ and\ \bibinfo {author}
  {\bibfnamefont {Morgan~W.}\ \bibnamefont {Mitchell}},\ }\bibfield  {title}
  {\enquote {\bibinfo {title} {{Improving Short-Term Stability in Optical
  Lattice Clocks by Quantum Nondemolition Measurements}},}\ }\href {\doibase
  10.1103/PhysRevLett.128.153201} {\bibfield  {journal} {\bibinfo  {journal}
  {Phys. Rev. Lett.}\ }\textbf {\bibinfo {volume} {128}},\ \bibinfo {pages}
  {153201} (\bibinfo {year} {2021})}\BibitemShut {NoStop}%
\bibitem [{\citenamefont {Baumgart}\ \emph {et~al.}(2016)\citenamefont
  {Baumgart}, \citenamefont {Cai}, \citenamefont {Retzker}, \citenamefont
  {Plenio},\ and\ \citenamefont {Wunderlich}}]{Baumgart2016}%
  \BibitemOpen
  \bibfield  {author} {\bibinfo {author} {\bibfnamefont {I.}~\bibnamefont
  {Baumgart}}, \bibinfo {author} {\bibfnamefont {J.-M.}\ \bibnamefont {Cai}},
  \bibinfo {author} {\bibfnamefont {A.}~\bibnamefont {Retzker}}, \bibinfo
  {author} {\bibfnamefont {M.~B.}\ \bibnamefont {Plenio}}, \ and\ \bibinfo
  {author} {\bibfnamefont {Ch}~\bibnamefont {Wunderlich}},\ }\bibfield  {title}
  {\enquote {\bibinfo {title} {{Ultrasensitive Magnetometer using a Single
  Atom}},}\ }\href {\doibase 10.1103/PhysRevLett.116.240801} {\bibfield
  {journal} {\bibinfo  {journal} {Phys. Rev. Lett.}\ }\textbf {\bibinfo
  {volume} {116}},\ \bibinfo {pages} {240801} (\bibinfo {year}
  {2016})}\BibitemShut {NoStop}%
\bibitem [{\citenamefont {Smullin}\ \emph {et~al.}(2009)\citenamefont
  {Smullin}, \citenamefont {Savukov}, \citenamefont {Vasilakis}, \citenamefont
  {Ghosh},\ and\ \citenamefont {Romalis}}]{PhysRevA.80.033420}%
  \BibitemOpen
  \bibfield  {author} {\bibinfo {author} {\bibfnamefont {S.~J.}\ \bibnamefont
  {Smullin}}, \bibinfo {author} {\bibfnamefont {I.~M.}\ \bibnamefont
  {Savukov}}, \bibinfo {author} {\bibfnamefont {G.}~\bibnamefont {Vasilakis}},
  \bibinfo {author} {\bibfnamefont {R.~K.}\ \bibnamefont {Ghosh}}, \ and\
  \bibinfo {author} {\bibfnamefont {M.~V.}\ \bibnamefont {Romalis}},\
  }\bibfield  {title} {\enquote {\bibinfo {title} {Low-noise high-density
  alkali-metal scalar magnetometer},}\ }\href {\doibase
  10.1103/PhysRevA.80.033420} {\bibfield  {journal} {\bibinfo  {journal} {Phys.
  Rev. A}\ }\textbf {\bibinfo {volume} {80}},\ \bibinfo {pages} {033420}
  (\bibinfo {year} {2009})}\BibitemShut {NoStop}%
\bibitem [{\citenamefont {Shi}\ \emph {et~al.}(2018)\citenamefont {Shi},
  \citenamefont {Jie}, \citenamefont {Li}, \citenamefont {Liu}, \citenamefont
  {Li},\ and\ \citenamefont {Zhang}}]{Hao2018}%
  \BibitemOpen
  \bibfield  {author} {\bibinfo {author} {\bibfnamefont {Hao}\ \bibnamefont
  {Shi}}, \bibinfo {author} {\bibfnamefont {Ma}~\bibnamefont {Jie}}, \bibinfo
  {author} {\bibfnamefont {Xiaofeng}\ \bibnamefont {Li}}, \bibinfo {author}
  {\bibfnamefont {Jie}\ \bibnamefont {Liu}}, \bibinfo {author} {\bibfnamefont
  {Chao}\ \bibnamefont {Li}}, \ and\ \bibinfo {author} {\bibfnamefont
  {Shougang}\ \bibnamefont {Zhang}},\ }\bibfield  {title} {\enquote {\bibinfo
  {title} {A quantum-based microwave magnetic field sensor},}\ }\href {\doibase
  10.3390/s18103288} {\bibfield  {journal} {\bibinfo  {journal} {Sensors}\
  }\textbf {\bibinfo {volume} {18}},\ \bibinfo {pages} {3288} (\bibinfo {year}
  {2018})}\BibitemShut {NoStop}%
\bibitem [{\citenamefont {Simpson}\ \emph {et~al.}(1963)\citenamefont
  {Simpson}, \citenamefont {Fraser},\ and\ \citenamefont
  {Greenwood}}]{4319483}%
  \BibitemOpen
  \bibfield  {author} {\bibinfo {author} {\bibfnamefont {J.~H.}\ \bibnamefont
  {Simpson}}, \bibinfo {author} {\bibfnamefont {J.~T.}\ \bibnamefont {Fraser}},
  \ and\ \bibinfo {author} {\bibfnamefont {I.~A.}\ \bibnamefont {Greenwood}},\
  }\bibfield  {title} {\enquote {\bibinfo {title} {An optically pumped nuclear
  magnetic resonance gyroscope},}\ }\href {\doibase 10.1109/TA.1963.4319483}
  {\bibfield  {journal} {\bibinfo  {journal} {IEEE Transactions on Aerospace}\
  }\textbf {\bibinfo {volume} {1}},\ \bibinfo {pages} {1107--1110} (\bibinfo
  {year} {1963})}\BibitemShut {NoStop}%
\bibitem [{\citenamefont {Szigeti}\ \emph {et~al.}(2021)\citenamefont
  {Szigeti}, \citenamefont {Hosten},\ and\ \citenamefont
  {Haine}}]{Szigeti2021}%
  \BibitemOpen
  \bibfield  {author} {\bibinfo {author} {\bibfnamefont {Stuart~S}\
  \bibnamefont {Szigeti}}, \bibinfo {author} {\bibfnamefont {Onur}\
  \bibnamefont {Hosten}}, \ and\ \bibinfo {author} {\bibfnamefont {Simon~A.}\
  \bibnamefont {Haine}},\ }\bibfield  {title} {\enquote {\bibinfo {title}
  {Improving cold-atom sensors with quantum entanglement: Prospects and
  challenges},}\ }\href {\doibase 10.1063/5.0050235} {\bibfield  {journal}
  {\bibinfo  {journal} {Applied Physics Letters}\ }\textbf {\bibinfo {volume}
  {118}},\ \bibinfo {pages} {140501} (\bibinfo {year} {2021})}\BibitemShut
  {NoStop}%
\bibitem [{\citenamefont {Giovannetti}\ \emph {et~al.}(2006)\citenamefont
  {Giovannetti}, \citenamefont {Lloyd},\ and\ \citenamefont
  {Maccone}}]{Giovannetti2006}%
  \BibitemOpen
  \bibfield  {author} {\bibinfo {author} {\bibfnamefont {Vittorio}\
  \bibnamefont {Giovannetti}}, \bibinfo {author} {\bibfnamefont {Seth}\
  \bibnamefont {Lloyd}}, \ and\ \bibinfo {author} {\bibfnamefont {Lorenzo}\
  \bibnamefont {Maccone}},\ }\bibfield  {title} {\enquote {\bibinfo {title}
  {Quantum metrology},}\ }\href@noop {} {\bibfield  {journal} {\bibinfo
  {journal} {Phys. Rev. Lett.}\ }\textbf {\bibinfo {volume} {96}},\ \bibinfo
  {pages} {010401} (\bibinfo {year} {2006})}\BibitemShut {NoStop}%
\bibitem [{\citenamefont {Lee}(2006)}]{Lee2006}%
  \BibitemOpen
  \bibfield  {author} {\bibinfo {author} {\bibfnamefont {Chaohong}\
  \bibnamefont {Lee}},\ }\bibfield  {title} {\enquote {\bibinfo {title}
  {Adiabatic mach-zehnder interferometry on a quantized bose-josephson
  junction},}\ }\href {\doibase 10.1103/PhysRevLett.97.150402} {\bibfield
  {journal} {\bibinfo  {journal} {Phys. Rev. Lett.}\ }\textbf {\bibinfo
  {volume} {97}},\ \bibinfo {pages} {150402} (\bibinfo {year}
  {2006})}\BibitemShut {NoStop}%
\bibitem [{\citenamefont {Huang}\ \emph {et~al.}(2014)\citenamefont {Huang},
  \citenamefont {Wu}, \citenamefont {Zhong},\ and\ \citenamefont
  {Lee}}]{Huang2014}%
  \BibitemOpen
  \bibfield  {author} {\bibinfo {author} {\bibfnamefont {Jiahao}\ \bibnamefont
  {Huang}}, \bibinfo {author} {\bibfnamefont {Shuyuan}\ \bibnamefont {Wu}},
  \bibinfo {author} {\bibfnamefont {Honghua}\ \bibnamefont {Zhong}}, \ and\
  \bibinfo {author} {\bibfnamefont {Chaohong}\ \bibnamefont {Lee}},\
  }\href@noop {} {\emph {\bibinfo {title} {Quantum Metrology with Cold
  Atoms}}},\ Vol.~\bibinfo {volume} {2}\ (\bibinfo {year} {2014})\ pp.\
  \bibinfo {pages} {365--415}\BibitemShut {NoStop}%
\bibitem [{\citenamefont {Pezz{\`e}}\ \emph {et~al.}(2018)\citenamefont
  {Pezz{\`e}}, \citenamefont {Smerzi}, \citenamefont {Oberthaler},
  \citenamefont {Schmied},\ and\ \citenamefont {Treutlein}}]{Pezze2018}%
  \BibitemOpen
  \bibfield  {author} {\bibinfo {author} {\bibfnamefont {Luca}\ \bibnamefont
  {Pezz{\`e}}}, \bibinfo {author} {\bibfnamefont {Augusto}\ \bibnamefont
  {Smerzi}}, \bibinfo {author} {\bibfnamefont {Markus~K.}\ \bibnamefont
  {Oberthaler}}, \bibinfo {author} {\bibfnamefont {Roman}\ \bibnamefont
  {Schmied}}, \ and\ \bibinfo {author} {\bibfnamefont {Philipp}\ \bibnamefont
  {Treutlein}},\ }\bibfield  {title} {\enquote {\bibinfo {title} {Quantum
  metrology with nonclassical states of atomic ensembles},}\ }\href {\doibase
  10.1103/RevModPhys.90.035005} {\bibfield  {journal} {\bibinfo  {journal}
  {Rev. Mod. Phys.}\ }\textbf {\bibinfo {volume} {90}},\ \bibinfo {pages}
  {035005} (\bibinfo {year} {2018})}\BibitemShut {NoStop}%
\bibitem [{\citenamefont {Lu}\ \emph {et~al.}(2019)\citenamefont {Lu},
  \citenamefont {Han}, \citenamefont {Zhuang}, \citenamefont {Ke},
  \citenamefont {Huang},\ and\ \citenamefont {Lee}}]{Lu2019}%
  \BibitemOpen
  \bibfield  {author} {\bibinfo {author} {\bibfnamefont {Bo}~\bibnamefont
  {Lu}}, \bibinfo {author} {\bibfnamefont {Cheng-Yin}\ \bibnamefont {Han}},
  \bibinfo {author} {\bibfnamefont {Min}\ \bibnamefont {Zhuang}}, \bibinfo
  {author} {\bibfnamefont {Yong-Guan}\ \bibnamefont {Ke}}, \bibinfo {author}
  {\bibfnamefont {Jia-Hao}\ \bibnamefont {Huang}}, \ and\ \bibinfo {author}
  {\bibfnamefont {Chao-Hong}\ \bibnamefont {Lee}},\ }\bibfield  {title}
  {\enquote {\bibinfo {title} {Non-gaussian entangled states and quantum
  metrology with ultracold atomic ensemble},}\ }\href {\doibase
  10.7498/aps.68.20190147} {\bibfield  {journal} {\bibinfo  {journal} {Acta
  Physica Sinica}\ }\textbf {\bibinfo {volume} {68}},\ \bibinfo {pages}
  {040306} (\bibinfo {year} {2019})}\BibitemShut {NoStop}%
\bibitem [{\citenamefont {Rabi}\ \emph {et~al.}(1938)\citenamefont {Rabi},
  \citenamefont {Millman}, \citenamefont {Kusch},\ and\ \citenamefont
  {Zacharias}}]{Rabi1938}%
  \BibitemOpen
  \bibfield  {author} {\bibinfo {author} {\bibfnamefont {I~I}\ \bibnamefont
  {Rabi}}, \bibinfo {author} {\bibfnamefont {S}~\bibnamefont {Millman}},
  \bibinfo {author} {\bibfnamefont {P}~\bibnamefont {Kusch}}, \ and\ \bibinfo
  {author} {\bibfnamefont {J~R}\ \bibnamefont {Zacharias}},\ }\bibfield
  {title} {\enquote {\bibinfo {title} {The molecular beam resonance method for
  measuring nuclear magnetic moments},}\ }\href@noop {} {\bibfield  {journal}
  {\bibinfo  {journal} {Phys. Rev}\ }\textbf {\bibinfo {volume} {53}},\
  \bibinfo {pages} {318} (\bibinfo {year} {1938})}\BibitemShut {NoStop}%
\bibitem [{\citenamefont {Ramsey}(1949)}]{Ramsey1949}%
  \BibitemOpen
  \bibfield  {author} {\bibinfo {author} {\bibfnamefont {Norman~F.}\
  \bibnamefont {Ramsey}},\ }\bibfield  {title} {\enquote {\bibinfo {title} {A
  new molecular beam resonance method},}\ }\href {\doibase
  10.1103/PhysRev.76.996} {\bibfield  {journal} {\bibinfo  {journal} {Physical
  Review}\ }\textbf {\bibinfo {volume} {76}},\ \bibinfo {pages} {996--996}
  (\bibinfo {year} {1949})}\BibitemShut {NoStop}%
\bibitem [{\citenamefont {Ramsey}(1950)}]{Ramsey1950}%
  \BibitemOpen
  \bibfield  {author} {\bibinfo {author} {\bibfnamefont {Norman~F.}\
  \bibnamefont {Ramsey}},\ }\bibfield  {title} {\enquote {\bibinfo {title} {A
  molecular beam resonance method with separated oscillating fields},}\ }\href
  {\doibase 10.1103/PhysRev.78.695} {\bibfield  {journal} {\bibinfo  {journal}
  {Physical Review}\ }\textbf {\bibinfo {volume} {78}},\ \bibinfo {pages}
  {695--699} (\bibinfo {year} {1950})}\BibitemShut {NoStop}%
\bibitem [{\citenamefont {Ramsey}(1980)}]{Ramsey1980}%
  \BibitemOpen
  \bibfield  {author} {\bibinfo {author} {\bibfnamefont {Norman~F.}\
  \bibnamefont {Ramsey}},\ }\bibfield  {title} {\enquote {\bibinfo {title} {The
  method of successive oscillatory fields},}\ }\href {\doibase
  10.1063/1.2914161} {\bibfield  {journal} {\bibinfo  {journal} {Physics
  Today}\ }\textbf {\bibinfo {volume} {33}},\ \bibinfo {pages} {25--30}
  (\bibinfo {year} {1980})}\BibitemShut {NoStop}%
\bibitem [{\citenamefont {Ludlow}\ \emph {et~al.}(2015)\citenamefont {Ludlow},
  \citenamefont {Boyd}, \citenamefont {Ye}, \citenamefont {Peik},\ and\
  \citenamefont {Schmidt}}]{Ludlow2015}%
  \BibitemOpen
  \bibfield  {author} {\bibinfo {author} {\bibfnamefont {Andrew~D}\
  \bibnamefont {Ludlow}}, \bibinfo {author} {\bibfnamefont {Martin~M}\
  \bibnamefont {Boyd}}, \bibinfo {author} {\bibfnamefont {Jun}\ \bibnamefont
  {Ye}}, \bibinfo {author} {\bibfnamefont {Ekkehard}\ \bibnamefont {Peik}}, \
  and\ \bibinfo {author} {\bibfnamefont {P.~O.}\ \bibnamefont {Schmidt}},\
  }\bibfield  {title} {\enquote {\bibinfo {title} {{Optical atomic clocks}},}\
  }\href {\doibase 10.1103/RevModPhys.87.637} {\bibfield  {journal} {\bibinfo
  {journal} {Rev. Mod. Phys.}\ }\textbf {\bibinfo {volume} {87}},\ \bibinfo
  {pages} {637--701} (\bibinfo {year} {2015})}\BibitemShut {NoStop}%
\bibitem [{\citenamefont {Kokkelmans}\ \emph {et~al.}(1997)\citenamefont
  {Kokkelmans}, \citenamefont {Verhaar}, \citenamefont {Gibble},\ and\
  \citenamefont {Heinzen}}]{PhysRevA.56.R4389}%
  \BibitemOpen
  \bibfield  {author} {\bibinfo {author} {\bibfnamefont {S.~J. J. M.~F.}\
  \bibnamefont {Kokkelmans}}, \bibinfo {author} {\bibfnamefont {B.~J.}\
  \bibnamefont {Verhaar}}, \bibinfo {author} {\bibfnamefont {K.}~\bibnamefont
  {Gibble}}, \ and\ \bibinfo {author} {\bibfnamefont {D.~J.}\ \bibnamefont
  {Heinzen}},\ }\bibfield  {title} {\enquote {\bibinfo {title} {Predictions for
  laser-cooled rb clocks},}\ }\href {\doibase 10.1103/PhysRevA.56.R4389}
  {\bibfield  {journal} {\bibinfo  {journal} {Phys. Rev. A}\ }\textbf {\bibinfo
  {volume} {56}},\ \bibinfo {pages} {R4389--R4392} (\bibinfo {year}
  {1997})}\BibitemShut {NoStop}%
\bibitem [{\citenamefont {Szymaniec}\ \emph {et~al.}(2007)\citenamefont
  {Szymaniec}, \citenamefont {Cha{\l}upczak}, \citenamefont {Tiesinga},
  \citenamefont {Williams}, \citenamefont {Weyers},\ and\ \citenamefont
  {Wynands}}]{Szymaniec2007}%
  \BibitemOpen
  \bibfield  {author} {\bibinfo {author} {\bibfnamefont {K.}~\bibnamefont
  {Szymaniec}}, \bibinfo {author} {\bibfnamefont {W.}~\bibnamefont
  {Cha{\l}upczak}}, \bibinfo {author} {\bibfnamefont {E.}~\bibnamefont
  {Tiesinga}}, \bibinfo {author} {\bibfnamefont {C.~J.}\ \bibnamefont
  {Williams}}, \bibinfo {author} {\bibfnamefont {S.}~\bibnamefont {Weyers}}, \
  and\ \bibinfo {author} {\bibfnamefont {R.}~\bibnamefont {Wynands}},\
  }\bibfield  {title} {\enquote {\bibinfo {title} {{Cancellation of the
  Collisional Frequency Shift in Caesium Fountain Clocks}},}\ }\href {\doibase
  10.1103/PhysRevLett.98.153002} {\bibfield  {journal} {\bibinfo  {journal}
  {Phys. Rev. Lett.}\ }\textbf {\bibinfo {volume} {98}},\ \bibinfo {pages}
  {153002} (\bibinfo {year} {2007})}\BibitemShut {NoStop}%
\bibitem [{\citenamefont {Yu}\ and\ \citenamefont
  {Pethick}(2010)}]{PhysRevLett.104.010801}%
  \BibitemOpen
  \bibfield  {author} {\bibinfo {author} {\bibfnamefont {Zhenhua}\ \bibnamefont
  {Yu}}\ and\ \bibinfo {author} {\bibfnamefont {C.~J.}\ \bibnamefont
  {Pethick}},\ }\bibfield  {title} {\enquote {\bibinfo {title} {Clock shifts of
  optical transitions in ultracold atomic gases},}\ }\href {\doibase
  10.1103/PhysRevLett.104.010801} {\bibfield  {journal} {\bibinfo  {journal}
  {Phys. Rev. Lett.}\ }\textbf {\bibinfo {volume} {104}},\ \bibinfo {pages}
  {010801} (\bibinfo {year} {2010})}\BibitemShut {NoStop}%
\bibitem [{\citenamefont {Lee}\ \emph {et~al.}(2016)\citenamefont {Lee},
  \citenamefont {Park}, \citenamefont {Lee},\ and\ \citenamefont
  {Yu}}]{Lee2016}%
  \BibitemOpen
  \bibfield  {author} {\bibinfo {author} {\bibfnamefont {Sangkyung}\
  \bibnamefont {Lee}}, \bibinfo {author} {\bibfnamefont {Chang~Yong}\
  \bibnamefont {Park}}, \bibinfo {author} {\bibfnamefont {Won~Kyu}\
  \bibnamefont {Lee}}, \ and\ \bibinfo {author} {\bibfnamefont {Dai~Hyuk}\
  \bibnamefont {Yu}},\ }\bibfield  {title} {\enquote {\bibinfo {title}
  {{Cancellation of collisional frequency shifts in optical lattice clocks with
  Rabi spectroscopy}},}\ }\href@noop {} {\bibfield  {journal} {\bibinfo
  {journal} {New Journal of Physics}\ }\textbf {\bibinfo {volume} {18}}
  (\bibinfo {year} {2016})}\BibitemShut {NoStop}%
\bibitem [{\citenamefont {Shaniv}\ \emph {et~al.}(2018)\citenamefont {Shaniv},
  \citenamefont {Manovitz}, \citenamefont {Shapira}, \citenamefont {Akerman},\
  and\ \citenamefont {Ozeri}}]{PhysRevLett.120.243603}%
  \BibitemOpen
  \bibfield  {author} {\bibinfo {author} {\bibfnamefont {Ravid}\ \bibnamefont
  {Shaniv}}, \bibinfo {author} {\bibfnamefont {Tom}\ \bibnamefont {Manovitz}},
  \bibinfo {author} {\bibfnamefont {Yotam}\ \bibnamefont {Shapira}}, \bibinfo
  {author} {\bibfnamefont {Nitzan}\ \bibnamefont {Akerman}}, \ and\ \bibinfo
  {author} {\bibfnamefont {Roee}\ \bibnamefont {Ozeri}},\ }\bibfield  {title}
  {\enquote {\bibinfo {title} {Toward heisenberg-limited rabi spectroscopy},}\
  }\href {\doibase 10.1103/PhysRevLett.120.243603} {\bibfield  {journal}
  {\bibinfo  {journal} {Phys. Rev. Lett.}\ }\textbf {\bibinfo {volume} {120}},\
  \bibinfo {pages} {243603} (\bibinfo {year} {2018})}\BibitemShut {NoStop}%
\bibitem [{SM()}]{SM}%
  \BibitemOpen
  \href@noop {} {}\bibinfo {note} {See Supplemental Material for details on:
  (i) Derivation of the half population difference Eq. (4), (ii) Measurement
  precisions for noninteracting systems, (iii) Analytical analysis on
  antisymmetric locking signal for noninteracting systems, (iv) Analytical
  analysis on antisymmetric locking signal for interacting systems, (v)
  Influences of imperfect $\pi/2$ pulse for preparing the initial state, (vi)
  Experimental feasibility via Bose condensed atoms, and (vii) Experimental
  feasibility via collective Cavity-QED system.}\BibitemShut {Stop}%
\bibitem [{\citenamefont {Sanner}\ \emph {et~al.}(2018)\citenamefont {Sanner},
  \citenamefont {Huntemann}, \citenamefont {Lange}, \citenamefont {Tamm},\ and\
  \citenamefont {Peik}}]{Sanner2018}%
  \BibitemOpen
  \bibfield  {author} {\bibinfo {author} {\bibfnamefont {Christian}\
  \bibnamefont {Sanner}}, \bibinfo {author} {\bibfnamefont {Nils}\ \bibnamefont
  {Huntemann}}, \bibinfo {author} {\bibfnamefont {Richard}\ \bibnamefont
  {Lange}}, \bibinfo {author} {\bibfnamefont {Christian}\ \bibnamefont {Tamm}},
  \ and\ \bibinfo {author} {\bibfnamefont {Ekkehard}\ \bibnamefont {Peik}},\
  }\bibfield  {title} {\enquote {\bibinfo {title} {Autobalanced ramsey
  spectroscopy},}\ }\href {\doibase 10.1103/PhysRevLett.120.053602} {\bibfield
  {journal} {\bibinfo  {journal} {Phys. Rev. Lett.}\ }\textbf {\bibinfo
  {volume} {120}},\ \bibinfo {pages} {53602} (\bibinfo {year}
  {2018})}\BibitemShut {NoStop}%
\bibitem [{\citenamefont {Lee}(2009)}]{PhysRevLett.102.070401}%
  \BibitemOpen
  \bibfield  {author} {\bibinfo {author} {\bibfnamefont {Chaohong}\
  \bibnamefont {Lee}},\ }\bibfield  {title} {\enquote {\bibinfo {title}
  {Universality and anomalous mean-field breakdown of symmetry-breaking
  transitions in a coupled two-component bose-einstein condensate},}\ }\href
  {\doibase 10.1103/PhysRevLett.102.070401} {\bibfield  {journal} {\bibinfo
  {journal} {Phys. Rev. Lett.}\ }\textbf {\bibinfo {volume} {102}},\ \bibinfo
  {pages} {070401} (\bibinfo {year} {2009})}\BibitemShut {NoStop}%
\bibitem [{\citenamefont {Trenkwalder}\ \emph {et~al.}(2016)\citenamefont
  {Trenkwalder}, \citenamefont {Spagnolli}, \citenamefont {Semeghini},
  \citenamefont {Coop}, \citenamefont {Landini}, \citenamefont {Castilho},
  \citenamefont {Pezz{\`e}}, \citenamefont {Modugno}, \citenamefont {Inguscio},
  \citenamefont {Smerzi},\ and\ \citenamefont {Fattori}}]{Trenkwalder2016}%
  \BibitemOpen
  \bibfield  {author} {\bibinfo {author} {\bibfnamefont {A.}~\bibnamefont
  {Trenkwalder}}, \bibinfo {author} {\bibfnamefont {G.}~\bibnamefont
  {Spagnolli}}, \bibinfo {author} {\bibfnamefont {G.}~\bibnamefont
  {Semeghini}}, \bibinfo {author} {\bibfnamefont {S.}~\bibnamefont {Coop}},
  \bibinfo {author} {\bibfnamefont {M.}~\bibnamefont {Landini}}, \bibinfo
  {author} {\bibfnamefont {P.}~\bibnamefont {Castilho}}, \bibinfo {author}
  {\bibfnamefont {L.}~\bibnamefont {Pezz{\`e}}}, \bibinfo {author}
  {\bibfnamefont {G.}~\bibnamefont {Modugno}}, \bibinfo {author} {\bibfnamefont
  {M.}~\bibnamefont {Inguscio}}, \bibinfo {author} {\bibfnamefont
  {A.}~\bibnamefont {Smerzi}}, \ and\ \bibinfo {author} {\bibfnamefont
  {M.}~\bibnamefont {Fattori}},\ }\bibfield  {title} {\enquote {\bibinfo
  {title} {Quantum phase transitions with parity-symmetry breaking and
  hysteresis},}\ }\href {\doibase 10.1038/nphys3743} {\bibfield  {journal}
  {\bibinfo  {journal} {Nature Physics}\ }\textbf {\bibinfo {volume} {12}},\
  \bibinfo {pages} {826--829} (\bibinfo {year} {2016})}\BibitemShut {NoStop}%
\bibitem [{\citenamefont {Zhuang}\ \emph {et~al.}(2020)\citenamefont {Zhuang},
  \citenamefont {Huang}, \citenamefont {Ke},\ and\ \citenamefont
  {Lee}}]{Zhuang2020}%
  \BibitemOpen
  \bibfield  {author} {\bibinfo {author} {\bibfnamefont {Min}\ \bibnamefont
  {Zhuang}}, \bibinfo {author} {\bibfnamefont {Jiahao}\ \bibnamefont {Huang}},
  \bibinfo {author} {\bibfnamefont {Yongguan}\ \bibnamefont {Ke}}, \ and\
  \bibinfo {author} {\bibfnamefont {Chaohong}\ \bibnamefont {Lee}},\ }\bibfield
   {title} {\enquote {\bibinfo {title} {Symmetry-protected quantum adiabatic
  evolution in spontaneous symmetry-breaking transitions},}\ }\href {\doibase
  10.1002/andp.201900471} {\bibfield  {journal} {\bibinfo  {journal} {Annalen
  der Physik}\ }\textbf {\bibinfo {volume} {532}},\ \bibinfo {pages} {1900471}
  (\bibinfo {year} {2020})}\BibitemShut {NoStop}%
\bibitem [{\citenamefont {Takamoto}\ \emph {et~al.}(2005)\citenamefont
  {Takamoto}, \citenamefont {Hong}, \citenamefont {Higashi},\ and\
  \citenamefont {Katori}}]{Takamoto2005}%
  \BibitemOpen
  \bibfield  {author} {\bibinfo {author} {\bibfnamefont {Masao}\ \bibnamefont
  {Takamoto}}, \bibinfo {author} {\bibfnamefont {Feng-Lei}\ \bibnamefont
  {Hong}}, \bibinfo {author} {\bibfnamefont {Ryoichi}\ \bibnamefont {Higashi}},
  \ and\ \bibinfo {author} {\bibfnamefont {Hidetoshi}\ \bibnamefont {Katori}},\
  }\bibfield  {title} {\enquote {\bibinfo {title} {{An optical lattice
  clock}},}\ }\href {\doibase 10.1038/nature03541} {\bibfield  {journal}
  {\bibinfo  {journal} {Nature}\ }\textbf {\bibinfo {volume} {435}},\ \bibinfo
  {pages} {321--324} (\bibinfo {year} {2005})}\BibitemShut {NoStop}%
\bibitem [{\citenamefont {Norcia}\ \emph {et~al.}(2019)\citenamefont {Norcia},
  \citenamefont {Young}, \citenamefont {Eckner}, \citenamefont {Oelker},
  \citenamefont {Ye},\ and\ \citenamefont {Kaufman}}]{Norcia2019}%
  \BibitemOpen
  \bibfield  {author} {\bibinfo {author} {\bibfnamefont {Matthew~A.}\
  \bibnamefont {Norcia}}, \bibinfo {author} {\bibfnamefont {Aaron~W.}\
  \bibnamefont {Young}}, \bibinfo {author} {\bibfnamefont {William~J.}\
  \bibnamefont {Eckner}}, \bibinfo {author} {\bibfnamefont {Eric}\ \bibnamefont
  {Oelker}}, \bibinfo {author} {\bibfnamefont {Jun}\ \bibnamefont {Ye}}, \ and\
  \bibinfo {author} {\bibfnamefont {Adam~M.}\ \bibnamefont {Kaufman}},\
  }\bibfield  {title} {\enquote {\bibinfo {title} {{Seconds-scale coherence on
  an optical clock transition in a tweezer array}},}\ }\href {\doibase
  10.1126/science.aay0644} {\bibfield  {journal} {\bibinfo  {journal}
  {Science}\ }\textbf {\bibinfo {volume} {366}},\ \bibinfo {pages} {93--97}
  (\bibinfo {year} {2019})}\BibitemShut {NoStop}%
\bibitem [{\citenamefont {Yin}\ \emph {et~al.}(2022)\citenamefont {Yin},
  \citenamefont {Lu}, \citenamefont {Li}, \citenamefont {Xia}, \citenamefont
  {Wang}, \citenamefont {Zhang},\ and\ \citenamefont
  {Chang}}]{PhysRevLett.128.073603}%
  \BibitemOpen
  \bibfield  {author} {\bibinfo {author} {\bibfnamefont {Mo-Juan}\ \bibnamefont
  {Yin}}, \bibinfo {author} {\bibfnamefont {Xiao-Tong}\ \bibnamefont {Lu}},
  \bibinfo {author} {\bibfnamefont {Ting}\ \bibnamefont {Li}}, \bibinfo
  {author} {\bibfnamefont {Jing-Jing}\ \bibnamefont {Xia}}, \bibinfo {author}
  {\bibfnamefont {Tao}\ \bibnamefont {Wang}}, \bibinfo {author} {\bibfnamefont
  {Xue-Feng}\ \bibnamefont {Zhang}}, \ and\ \bibinfo {author} {\bibfnamefont
  {Hong}\ \bibnamefont {Chang}},\ }\bibfield  {title} {\enquote {\bibinfo
  {title} {Floquet engineering hz-level rabi spectra in shallow optical lattice
  clock},}\ }\href {\doibase 10.1103/PhysRevLett.128.073603} {\bibfield
  {journal} {\bibinfo  {journal} {Phys. Rev. Lett.}\ }\textbf {\bibinfo
  {volume} {128}},\ \bibinfo {pages} {073603} (\bibinfo {year}
  {2022})}\BibitemShut {NoStop}%
\bibitem [{\citenamefont {Liu}\ \emph {et~al.}(2017)\citenamefont {Liu},
  \citenamefont {Zhang}, \citenamefont {Jiang}, \citenamefont {Wang},
  \citenamefont {Zhu}, \citenamefont {Xiong}, \citenamefont {He},\ and\
  \citenamefont {Lyu}}]{Liu_2017}%
  \BibitemOpen
  \bibfield  {author} {\bibinfo {author} {\bibfnamefont {Hui}\ \bibnamefont
  {Liu}}, \bibinfo {author} {\bibfnamefont {Xi}~\bibnamefont {Zhang}}, \bibinfo
  {author} {\bibfnamefont {Kun-Liang}\ \bibnamefont {Jiang}}, \bibinfo {author}
  {\bibfnamefont {Jin-Qi}\ \bibnamefont {Wang}}, \bibinfo {author}
  {\bibfnamefont {Qiang}\ \bibnamefont {Zhu}}, \bibinfo {author} {\bibfnamefont
  {Zhuan-Xian}\ \bibnamefont {Xiong}}, \bibinfo {author} {\bibfnamefont
  {Ling-Xiang}\ \bibnamefont {He}}, \ and\ \bibinfo {author} {\bibfnamefont
  {Bao-Long}\ \bibnamefont {Lyu}},\ }\bibfield  {title} {\enquote {\bibinfo
  {title} {Realization of closed-loop operation of optical lattice clock based
  on 171yb*},}\ }\href {\doibase 10.1088/0256-307X/34/2/020601} {\bibfield
  {journal} {\bibinfo  {journal} {Chinese Physics Letters}\ }\textbf {\bibinfo
  {volume} {34}},\ \bibinfo {pages} {020601} (\bibinfo {year}
  {2017})}\BibitemShut {NoStop}%
\bibitem [{\citenamefont {Luo}\ \emph {et~al.}(2020)\citenamefont {Luo},
  \citenamefont {Qiao}, \citenamefont {Ai}, \citenamefont {Zhou}, \citenamefont
  {Zhang}, \citenamefont {Zhang}, \citenamefont {Sun}, \citenamefont {Qi},
  \citenamefont {Peng}, \citenamefont {Jin}, \citenamefont {Fang},
  \citenamefont {Yang}, \citenamefont {Li}, \citenamefont {Liang},\ and\
  \citenamefont {Xu}}]{Luo_2020}%
  \BibitemOpen
  \bibfield  {author} {\bibinfo {author} {\bibfnamefont {Limeng}\ \bibnamefont
  {Luo}}, \bibinfo {author} {\bibfnamefont {Hao}\ \bibnamefont {Qiao}},
  \bibinfo {author} {\bibfnamefont {Di}~\bibnamefont {Ai}}, \bibinfo {author}
  {\bibfnamefont {Min}\ \bibnamefont {Zhou}}, \bibinfo {author} {\bibfnamefont
  {Shuang}\ \bibnamefont {Zhang}}, \bibinfo {author} {\bibfnamefont {Sheng}\
  \bibnamefont {Zhang}}, \bibinfo {author} {\bibfnamefont {Changyue}\
  \bibnamefont {Sun}}, \bibinfo {author} {\bibfnamefont {Qichao}\ \bibnamefont
  {Qi}}, \bibinfo {author} {\bibfnamefont {Chengquan}\ \bibnamefont {Peng}},
  \bibinfo {author} {\bibfnamefont {Taoyun}\ \bibnamefont {Jin}}, \bibinfo
  {author} {\bibfnamefont {Wei}\ \bibnamefont {Fang}}, \bibinfo {author}
  {\bibfnamefont {Zhiqiang}\ \bibnamefont {Yang}}, \bibinfo {author}
  {\bibfnamefont {Tianchu}\ \bibnamefont {Li}}, \bibinfo {author}
  {\bibfnamefont {Kun}\ \bibnamefont {Liang}}, \ and\ \bibinfo {author}
  {\bibfnamefont {Xinye}\ \bibnamefont {Xu}},\ }\bibfield  {title} {\enquote
  {\bibinfo {title} {Absolute frequency measurement of an yb optical clock at
  the 10-16 level using international atomic time},}\ }\href {\doibase
  10.1088/1681-7575/abb879} {\bibfield  {journal} {\bibinfo  {journal}
  {Metrologia}\ }\textbf {\bibinfo {volume} {57}},\ \bibinfo {pages} {065017}
  (\bibinfo {year} {2020})}\BibitemShut {NoStop}%
\bibitem [{\citenamefont {Nolan}\ \emph {et~al.}(2017)\citenamefont {Nolan},
  \citenamefont {Szigeti},\ and\ \citenamefont {Haine}}]{Nolan2017}%
  \BibitemOpen
  \bibfield  {author} {\bibinfo {author} {\bibfnamefont {Samuel~P.}\
  \bibnamefont {Nolan}}, \bibinfo {author} {\bibfnamefont {Stuart~S.}\
  \bibnamefont {Szigeti}}, \ and\ \bibinfo {author} {\bibfnamefont {Simon~A.}\
  \bibnamefont {Haine}},\ }\bibfield  {title} {\enquote {\bibinfo {title}
  {Optimal and robust quantum metrology using interaction-based readouts},}\
  }\href {\doibase 10.1103/PhysRevLett.119.193601} {\bibfield  {journal}
  {\bibinfo  {journal} {Phys. Rev. Lett.}\ }\textbf {\bibinfo {volume} {119}},\
  \bibinfo {pages} {193601} (\bibinfo {year} {2017})}\BibitemShut {NoStop}%
\bibitem [{\citenamefont {Haine}(2018)}]{PhysRevA.98.030303}%
  \BibitemOpen
  \bibfield  {author} {\bibinfo {author} {\bibfnamefont {Simon~A.}\
  \bibnamefont {Haine}},\ }\bibfield  {title} {\enquote {\bibinfo {title}
  {Using interaction-based readouts to approach the ultimate limit of
  detection-noise robustness for quantum-enhanced metrology in collective spin
  systems},}\ }\href {\doibase 10.1103/PhysRevA.98.030303} {\bibfield
  {journal} {\bibinfo  {journal} {Phys. Rev. A}\ }\textbf {\bibinfo {volume}
  {98}},\ \bibinfo {pages} {030303} (\bibinfo {year} {2018})}\BibitemShut
  {NoStop}%
\bibitem [{\citenamefont {Huang}\ \emph
  {et~al.}(2018{\natexlab{a}})\citenamefont {Huang}, \citenamefont {Zhuang},\
  and\ \citenamefont {Lee}}]{Huang2018-2}%
  \BibitemOpen
  \bibfield  {author} {\bibinfo {author} {\bibfnamefont {Jiahao}\ \bibnamefont
  {Huang}}, \bibinfo {author} {\bibfnamefont {Min}\ \bibnamefont {Zhuang}}, \
  and\ \bibinfo {author} {\bibfnamefont {Chaohong}\ \bibnamefont {Lee}},\
  }\bibfield  {title} {\enquote {\bibinfo {title} {Non-gaussian precision
  metrology via driving through quantum phase transitions},}\ }\href {\doibase
  10.1103/PhysRevA.97.032116} {\bibfield  {journal} {\bibinfo  {journal} {Phys.
  Rev. A}\ }\textbf {\bibinfo {volume} {97}},\ \bibinfo {pages} {032116}
  (\bibinfo {year} {2018}{\natexlab{a}})}\BibitemShut {NoStop}%
\bibitem [{\citenamefont {Wineland}\ \emph {et~al.}(1992)\citenamefont
  {Wineland}, \citenamefont {Bollinger}, \citenamefont {Itano}, \citenamefont
  {Moore},\ and\ \citenamefont {Heinzen}}]{Wineland1992}%
  \BibitemOpen
  \bibfield  {author} {\bibinfo {author} {\bibfnamefont {D.~J.}\ \bibnamefont
  {Wineland}}, \bibinfo {author} {\bibfnamefont {J.~J.}\ \bibnamefont
  {Bollinger}}, \bibinfo {author} {\bibfnamefont {W.~M.}\ \bibnamefont
  {Itano}}, \bibinfo {author} {\bibfnamefont {F.~L.}\ \bibnamefont {Moore}}, \
  and\ \bibinfo {author} {\bibfnamefont {D.~J.}\ \bibnamefont {Heinzen}},\
  }\bibfield  {title} {\enquote {\bibinfo {title} {Spin squeezing and reduced
  quantum noise in spectroscopy},}\ }\href {\doibase 10.1103/PhysRevA.46.R6797}
  {\bibfield  {journal} {\bibinfo  {journal} {Phys. Rev. A}\ }\textbf {\bibinfo
  {volume} {46}},\ \bibinfo {pages} {R6797--R6800} (\bibinfo {year}
  {1992})}\BibitemShut {NoStop}%
\bibitem [{\citenamefont {Wineland}\ \emph {et~al.}(1994)\citenamefont
  {Wineland}, \citenamefont {Bollinger}, \citenamefont {Itano},\ and\
  \citenamefont {Heinzen}}]{Wineland1994}%
  \BibitemOpen
  \bibfield  {author} {\bibinfo {author} {\bibfnamefont {D.~J.}\ \bibnamefont
  {Wineland}}, \bibinfo {author} {\bibfnamefont {J.~J.}\ \bibnamefont
  {Bollinger}}, \bibinfo {author} {\bibfnamefont {W.~M.}\ \bibnamefont
  {Itano}}, \ and\ \bibinfo {author} {\bibfnamefont {D.~J.}\ \bibnamefont
  {Heinzen}},\ }\bibfield  {title} {\enquote {\bibinfo {title} {Squeezed atomic
  states and projection noise in spectroscopy},}\ }\href {\doibase
  10.1103/PhysRevA.50.67} {\bibfield  {journal} {\bibinfo  {journal} {Phys.
  Rev. A}\ }\textbf {\bibinfo {volume} {50}},\ \bibinfo {pages} {67--88}
  (\bibinfo {year} {1994})}\BibitemShut {NoStop}%
\bibitem [{\citenamefont {Louchet-Chauvet}\ \emph {et~al.}(2010)\citenamefont
  {Louchet-Chauvet}, \citenamefont {Appel}, \citenamefont {Renema},
  \citenamefont {Oblak}, \citenamefont {Kjaergaard},\ and\ \citenamefont
  {Polzik}}]{Louchet-Chauvet2010}%
  \BibitemOpen
  \bibfield  {author} {\bibinfo {author} {\bibfnamefont {Anne}\ \bibnamefont
  {Louchet-Chauvet}}, \bibinfo {author} {\bibfnamefont {J{\"{u}}rgen}\
  \bibnamefont {Appel}}, \bibinfo {author} {\bibfnamefont {Jelmer~J.}\
  \bibnamefont {Renema}}, \bibinfo {author} {\bibfnamefont {Daniel}\
  \bibnamefont {Oblak}}, \bibinfo {author} {\bibfnamefont {Niels}\ \bibnamefont
  {Kjaergaard}}, \ and\ \bibinfo {author} {\bibfnamefont {Eugene~S.}\
  \bibnamefont {Polzik}},\ }\bibfield  {title} {\enquote {\bibinfo {title}
  {{Entanglement-assisted atomic clock beyond the projection noise limit}},}\
  }\href {\doibase 10.1088/1367-2630/12/6/065032} {\bibfield  {journal}
  {\bibinfo  {journal} {New Journal of Physics}\ }\textbf {\bibinfo {volume}
  {12}},\ \bibinfo {pages} {065032} (\bibinfo {year} {2010})}\BibitemShut
  {NoStop}%
\bibitem [{\citenamefont {Kitagawa}\ and\ \citenamefont
  {Ueda}(1993)}]{Kitagawa1993}%
  \BibitemOpen
  \bibfield  {author} {\bibinfo {author} {\bibfnamefont {Masahiro}\
  \bibnamefont {Kitagawa}}\ and\ \bibinfo {author} {\bibfnamefont {Masahito}\
  \bibnamefont {Ueda}},\ }\bibfield  {title} {\enquote {\bibinfo {title}
  {Squeezed spin states},}\ }\href {\doibase 10.1103/PhysRevA.47.5138}
  {\bibfield  {journal} {\bibinfo  {journal} {Phys. Rev. A}\ }\textbf {\bibinfo
  {volume} {47}},\ \bibinfo {pages} {5138--5143} (\bibinfo {year}
  {1993})}\BibitemShut {NoStop}%
\bibitem [{\citenamefont {Ma}\ \emph {et~al.}(2011)\citenamefont {Ma},
  \citenamefont {Wang}, \citenamefont {Sun},\ and\ \citenamefont
  {Nori}}]{Ma2011}%
  \BibitemOpen
  \bibfield  {author} {\bibinfo {author} {\bibfnamefont {Jian}\ \bibnamefont
  {Ma}}, \bibinfo {author} {\bibfnamefont {Xiaoguang}\ \bibnamefont {Wang}},
  \bibinfo {author} {\bibfnamefont {C.P.}\ \bibnamefont {Sun}}, \ and\ \bibinfo
  {author} {\bibfnamefont {Franco}\ \bibnamefont {Nori}},\ }\bibfield  {title}
  {\enquote {\bibinfo {title} {Quantum spin squeezing},}\ }\href {\doibase
  10.1016/j.physrep.2011.08.003} {\bibfield  {journal} {\bibinfo  {journal}
  {Physics Reports}\ }\textbf {\bibinfo {volume} {509}},\ \bibinfo {pages}
  {89--165} (\bibinfo {year} {2011})}\BibitemShut {NoStop}%
\bibitem [{\citenamefont {Gross}(2012)}]{Gross2012-1}%
  \BibitemOpen
  \bibfield  {author} {\bibinfo {author} {\bibfnamefont {Christian}\
  \bibnamefont {Gross}},\ }\bibfield  {title} {\enquote {\bibinfo {title} {Spin
  squeezing, entanglement and quantum metrology with bose-einstein
  condensates},}\ }\href {\doibase 10.1088/0953-4075/45/10/103001} {\bibfield
  {journal} {\bibinfo  {journal} {Journal of Physics B}\ }\textbf {\bibinfo
  {volume} {45}},\ \bibinfo {pages} {103001} (\bibinfo {year}
  {2012})}\BibitemShut {NoStop}%
\bibitem [{\citenamefont {Davis}\ \emph {et~al.}(2016)\citenamefont {Davis},
  \citenamefont {Bentsen},\ and\ \citenamefont {Schleier-Smith}}]{Davis2016}%
  \BibitemOpen
  \bibfield  {author} {\bibinfo {author} {\bibfnamefont {Emily}\ \bibnamefont
  {Davis}}, \bibinfo {author} {\bibfnamefont {Gregory}\ \bibnamefont
  {Bentsen}}, \ and\ \bibinfo {author} {\bibfnamefont {Monika}\ \bibnamefont
  {Schleier-Smith}},\ }\bibfield  {title} {\enquote {\bibinfo {title}
  {Approaching the heisenberg limit without single-particle detection},}\
  }\href {\doibase 10.1103/PhysRevLett.116.053601} {\bibfield  {journal}
  {\bibinfo  {journal} {Phys. Rev. Lett.}\ }\textbf {\bibinfo {volume} {116}},\
  \bibinfo {pages} {053601} (\bibinfo {year} {2016})}\BibitemShut {NoStop}%
\bibitem [{\citenamefont {Huang}\ \emph
  {et~al.}(2018{\natexlab{b}})\citenamefont {Huang}, \citenamefont {Zhuang},
  \citenamefont {Lu}, \citenamefont {Ke},\ and\ \citenamefont
  {Lee}}]{Huang2018-1}%
  \BibitemOpen
  \bibfield  {author} {\bibinfo {author} {\bibfnamefont {Jiahao}\ \bibnamefont
  {Huang}}, \bibinfo {author} {\bibfnamefont {Min}\ \bibnamefont {Zhuang}},
  \bibinfo {author} {\bibfnamefont {Bo}~\bibnamefont {Lu}}, \bibinfo {author}
  {\bibfnamefont {Yongguan}\ \bibnamefont {Ke}}, \ and\ \bibinfo {author}
  {\bibfnamefont {Chaohong}\ \bibnamefont {Lee}},\ }\bibfield  {title}
  {\enquote {\bibinfo {title} {Achieving heisenberg-limited metrology with spin
  cat states via interaction-based readout},}\ }\href {\doibase
  10.1103/PhysRevA.98.012129} {\bibfield  {journal} {\bibinfo  {journal} {Phys.
  Rev. A}\ }\textbf {\bibinfo {volume} {98}},\ \bibinfo {pages} {012129}
  (\bibinfo {year} {2018}{\natexlab{b}})}\BibitemShut {NoStop}%
\bibitem [{\citenamefont {Hosten}\ \emph {et~al.}(2016)\citenamefont {Hosten},
  \citenamefont {Engelsen}, \citenamefont {Krishnakumar},\ and\ \citenamefont
  {Kasevich}}]{Hosten2016-2}%
  \BibitemOpen
  \bibfield  {author} {\bibinfo {author} {\bibfnamefont {Onur}\ \bibnamefont
  {Hosten}}, \bibinfo {author} {\bibfnamefont {Nils~J.}\ \bibnamefont
  {Engelsen}}, \bibinfo {author} {\bibfnamefont {Rajiv}\ \bibnamefont
  {Krishnakumar}}, \ and\ \bibinfo {author} {\bibfnamefont {Mark~A.}\
  \bibnamefont {Kasevich}},\ }\bibfield  {title} {\enquote {\bibinfo {title}
  {{Measurement noise 100 times lower than the quantum-projection limit using
  entangled atoms}},}\ }\href {\doibase 10.1038/nature16176} {\bibfield
  {journal} {\bibinfo  {journal} {Nature}\ }\textbf {\bibinfo {volume} {529}},\
  \bibinfo {pages} {505--508} (\bibinfo {year} {2016})}\BibitemShut {NoStop}%
\bibitem [{\citenamefont {Li}\ \emph {et~al.}(2021)\citenamefont {Li},
  \citenamefont {da~Silva}, \citenamefont {Kain}, \citenamefont {Pati},
  \citenamefont {Tripathi},\ and\ \citenamefont {Shahriar}}]{Li2021-2}%
  \BibitemOpen
  \bibfield  {author} {\bibinfo {author} {\bibfnamefont {Jinyang}\ \bibnamefont
  {Li}}, \bibinfo {author} {\bibfnamefont {Greg{\'{o}}rio R.~M.}\ \bibnamefont
  {da~Silva}}, \bibinfo {author} {\bibfnamefont {Schuyler}\ \bibnamefont
  {Kain}}, \bibinfo {author} {\bibfnamefont {Gour}\ \bibnamefont {Pati}},
  \bibinfo {author} {\bibfnamefont {Renu}\ \bibnamefont {Tripathi}}, \ and\
  \bibinfo {author} {\bibfnamefont {Selim~M.}\ \bibnamefont {Shahriar}},\
  }\bibfield  {title} {\enquote {\bibinfo {title} {{Spin Squeezing Induced
  Enhancement of Sensitivity of an Atomic Clock using Coherent Population
  Trapping}},}\ }\href {http://arxiv.org/abs/2112.08013} {\  (\bibinfo {year}
  {2021})},\ \Eprint {http://arxiv.org/abs/2112.08013} {arXiv:2112.08013}
  \BibitemShut {NoStop}%
\bibitem [{\citenamefont {Gross}\ \emph {et~al.}(2010)\citenamefont {Gross},
  \citenamefont {Zibold}, \citenamefont {Nicklas}, \citenamefont {Est{\`e}ve},\
  and\ \citenamefont {Oberthaler}}]{Gross2010}%
  \BibitemOpen
  \bibfield  {author} {\bibinfo {author} {\bibfnamefont {C.}~\bibnamefont
  {Gross}}, \bibinfo {author} {\bibfnamefont {T.}~\bibnamefont {Zibold}},
  \bibinfo {author} {\bibfnamefont {E.}~\bibnamefont {Nicklas}}, \bibinfo
  {author} {\bibfnamefont {J.}~\bibnamefont {Est{\`e}ve}}, \ and\ \bibinfo
  {author} {\bibfnamefont {M.~K.}\ \bibnamefont {Oberthaler}},\ }\bibfield
  {title} {\enquote {\bibinfo {title} {Nonlinear atom interferometer surpasses
  classical precision limit},}\ }\href {\doibase 10.1038/nature08919}
  {\bibfield  {journal} {\bibinfo  {journal} {Nature}\ }\textbf {\bibinfo
  {volume} {464}},\ \bibinfo {pages} {1165--1169} (\bibinfo {year}
  {2010})}\BibitemShut {NoStop}%
\bibitem [{\citenamefont {Riedel}\ \emph {et~al.}(2010)\citenamefont {Riedel},
  \citenamefont {B{\"o}hi}, \citenamefont {Li}, \citenamefont {Signnsch},
  \citenamefont {Sinatra},\ and\ \citenamefont {Treutlein}}]{Riedel2010}%
  \BibitemOpen
  \bibfield  {author} {\bibinfo {author} {\bibfnamefont {Max~F.}\ \bibnamefont
  {Riedel}}, \bibinfo {author} {\bibfnamefont {Pascal}\ \bibnamefont
  {B{\"o}hi}}, \bibinfo {author} {\bibfnamefont {Yun}\ \bibnamefont {Li}},
  \bibinfo {author} {\bibfnamefont {Theodor W.~H{\"a}nsch}\ \bibnamefont
  {Signnsch}}, \bibinfo {author} {\bibfnamefont {Alice}\ \bibnamefont
  {Sinatra}}, \ and\ \bibinfo {author} {\bibfnamefont {Philipp}\ \bibnamefont
  {Treutlein}},\ }\bibfield  {title} {\enquote {\bibinfo {title}
  {Atom-chip-based generation of entanglement for quantum metrology},}\ }\href
  {\doibase 10.1038/nature08988} {\bibfield  {journal} {\bibinfo  {journal}
  {Nature}\ }\textbf {\bibinfo {volume} {464}},\ \bibinfo {pages} {1170--1173}
  (\bibinfo {year} {2010})}\BibitemShut {NoStop}%
\bibitem [{\citenamefont {Ockeloen}\ \emph {et~al.}(2013)\citenamefont
  {Ockeloen}, \citenamefont {Schmied}, \citenamefont {Riedel},\ and\
  \citenamefont {Treutlein}}]{Ockeloen2013}%
  \BibitemOpen
  \bibfield  {author} {\bibinfo {author} {\bibfnamefont {Caspar~F.}\
  \bibnamefont {Ockeloen}}, \bibinfo {author} {\bibfnamefont {Roman}\
  \bibnamefont {Schmied}}, \bibinfo {author} {\bibfnamefont {Max~F.}\
  \bibnamefont {Riedel}}, \ and\ \bibinfo {author} {\bibfnamefont {Philipp}\
  \bibnamefont {Treutlein}},\ }\bibfield  {title} {\enquote {\bibinfo {title}
  {Quantum metrology with a scanning probe atom interferometer},}\ }\href
  {\doibase 10.1103/PhysRevLett.111.143001} {\bibfield  {journal} {\bibinfo
  {journal} {Phys. Rev. Lett.}\ }\textbf {\bibinfo {volume} {111}},\ \bibinfo
  {pages} {143001} (\bibinfo {year} {2013})}\BibitemShut {NoStop}%
\bibitem [{\citenamefont {Strobel}\ \emph {et~al.}(2014)\citenamefont
  {Strobel}, \citenamefont {Muessel}, \citenamefont {Linnemann}, \citenamefont
  {Zibold}, \citenamefont {Hume}, \citenamefont {Pezze}, \citenamefont
  {Smerzi},\ and\ \citenamefont {Oberthaler}}]{Strobel2014}%
  \BibitemOpen
  \bibfield  {author} {\bibinfo {author} {\bibfnamefont {Helmut}\ \bibnamefont
  {Strobel}}, \bibinfo {author} {\bibfnamefont {Wolfgang}\ \bibnamefont
  {Muessel}}, \bibinfo {author} {\bibfnamefont {Daniel}\ \bibnamefont
  {Linnemann}}, \bibinfo {author} {\bibfnamefont {Tilman}\ \bibnamefont
  {Zibold}}, \bibinfo {author} {\bibfnamefont {David~B.}\ \bibnamefont {Hume}},
  \bibinfo {author} {\bibfnamefont {L.}~\bibnamefont {Pezze}}, \bibinfo
  {author} {\bibfnamefont {Augusto}\ \bibnamefont {Smerzi}}, \ and\ \bibinfo
  {author} {\bibfnamefont {Markus~K.}\ \bibnamefont {Oberthaler}},\ }\bibfield
  {title} {\enquote {\bibinfo {title} {Fisher information and entanglement of
  non-gaussian spin states},}\ }\href {\doibase 10.1126/science.1250147}
  {\bibfield  {journal} {\bibinfo  {journal} {Science}\ }\textbf {\bibinfo
  {volume} {345}},\ \bibinfo {pages} {424--427} (\bibinfo {year}
  {2014})}\BibitemShut {NoStop}%
\bibitem [{\citenamefont {Colombo}\ \emph {et~al.}(2022)\citenamefont
  {Colombo}, \citenamefont {Pedrozo-Pe{\~{n}}afiel}, \citenamefont
  {Adiyatullin}, \citenamefont {Li}, \citenamefont {Mendez}, \citenamefont
  {Shu},\ and\ \citenamefont {Vuleti{\'{c}}}}]{Colombo2022}%
  \BibitemOpen
  \bibfield  {author} {\bibinfo {author} {\bibfnamefont {Simone}\ \bibnamefont
  {Colombo}}, \bibinfo {author} {\bibfnamefont {Edwin}\ \bibnamefont
  {Pedrozo-Pe{\~{n}}afiel}}, \bibinfo {author} {\bibfnamefont {Albert~F.}\
  \bibnamefont {Adiyatullin}}, \bibinfo {author} {\bibfnamefont {Zeyang}\
  \bibnamefont {Li}}, \bibinfo {author} {\bibfnamefont {Enrique}\ \bibnamefont
  {Mendez}}, \bibinfo {author} {\bibfnamefont {Chi}\ \bibnamefont {Shu}}, \
  and\ \bibinfo {author} {\bibfnamefont {Vladan}\ \bibnamefont
  {Vuleti{\'{c}}}},\ }\bibfield  {title} {\enquote {\bibinfo {title}
  {{Time-reversal-based quantum metrology with many-body entangled states}},}\
  }\href {\doibase 10.1038/s41567-022-01653-5} {\bibfield  {journal} {\bibinfo
  {journal} {Nature Physics}\ } (\bibinfo {year} {2022}),\
  10.1038/s41567-022-01653-5}\BibitemShut {NoStop}%
\bibitem [{\citenamefont {Pedrozo-Pe{\~{n}}afiel}\ \emph
  {et~al.}(2020)\citenamefont {Pedrozo-Pe{\~{n}}afiel}, \citenamefont
  {Colombo}, \citenamefont {Shu}, \citenamefont {Adiyatullin}, \citenamefont
  {Li}, \citenamefont {Mendez}, \citenamefont {Braverman}, \citenamefont
  {Kawasaki}, \citenamefont {Akamatsu}, \citenamefont {Xiao},\ and\
  \citenamefont {Vuleti{\'{c}}}}]{Pedrozo-Penafiel2020}%
  \BibitemOpen
  \bibfield  {author} {\bibinfo {author} {\bibfnamefont {Edwin}\ \bibnamefont
  {Pedrozo-Pe{\~{n}}afiel}}, \bibinfo {author} {\bibfnamefont {Simone}\
  \bibnamefont {Colombo}}, \bibinfo {author} {\bibfnamefont {Chi}\ \bibnamefont
  {Shu}}, \bibinfo {author} {\bibfnamefont {Albert~F.}\ \bibnamefont
  {Adiyatullin}}, \bibinfo {author} {\bibfnamefont {Zeyang}\ \bibnamefont
  {Li}}, \bibinfo {author} {\bibfnamefont {Enrique}\ \bibnamefont {Mendez}},
  \bibinfo {author} {\bibfnamefont {Boris}\ \bibnamefont {Braverman}}, \bibinfo
  {author} {\bibfnamefont {Akio}\ \bibnamefont {Kawasaki}}, \bibinfo {author}
  {\bibfnamefont {Daisuke}\ \bibnamefont {Akamatsu}}, \bibinfo {author}
  {\bibfnamefont {Yanhong}\ \bibnamefont {Xiao}}, \ and\ \bibinfo {author}
  {\bibfnamefont {Vladan}\ \bibnamefont {Vuleti{\'{c}}}},\ }\bibfield  {title}
  {\enquote {\bibinfo {title} {{Entanglement on an optical atomic-clock
  transition}},}\ }\href {\doibase 10.1038/s41586-020-3006-1} {\bibfield
  {journal} {\bibinfo  {journal} {Nature}\ }\textbf {\bibinfo {volume} {588}},\
  \bibinfo {pages} {414--418} (\bibinfo {year} {2020})}\BibitemShut {NoStop}%
\bibitem [{\citenamefont {Li}\ \emph {et~al.}(2022)\citenamefont {Li},
  \citenamefont {Braverman}, \citenamefont {Colombo}, \citenamefont {Shu},
  \citenamefont {Kawasaki}, \citenamefont {Adiyatullin}, \citenamefont
  {Pedrozo-Pe\~nafiel}, \citenamefont {Mendez},\ and\ \citenamefont
  {Vuleti\ifmmode~\acute{c}\else \'{c}\fi{}}}]{PRXQuantum.3.020308}%
  \BibitemOpen
  \bibfield  {author} {\bibinfo {author} {\bibfnamefont {Zeyang}\ \bibnamefont
  {Li}}, \bibinfo {author} {\bibfnamefont {Boris}\ \bibnamefont {Braverman}},
  \bibinfo {author} {\bibfnamefont {Simone}\ \bibnamefont {Colombo}}, \bibinfo
  {author} {\bibfnamefont {Chi}\ \bibnamefont {Shu}}, \bibinfo {author}
  {\bibfnamefont {Akio}\ \bibnamefont {Kawasaki}}, \bibinfo {author}
  {\bibfnamefont {Albert~F.}\ \bibnamefont {Adiyatullin}}, \bibinfo {author}
  {\bibfnamefont {Edwin}\ \bibnamefont {Pedrozo-Pe\~nafiel}}, \bibinfo {author}
  {\bibfnamefont {Enrique}\ \bibnamefont {Mendez}}, \ and\ \bibinfo {author}
  {\bibfnamefont {Vladan}\ \bibnamefont {Vuleti\ifmmode~\acute{c}\else
  \'{c}\fi{}}},\ }\bibfield  {title} {\enquote {\bibinfo {title} {Collective
  spin-light and light-mediated spin-spin interactions in an optical cavity},}\
  }\href {\doibase 10.1103/PRXQuantum.3.020308} {\bibfield  {journal} {\bibinfo
   {journal} {PRX Quantum}\ }\textbf {\bibinfo {volume} {3}},\ \bibinfo {pages}
  {020308} (\bibinfo {year} {2022})}\BibitemShut {NoStop}%
\bibitem [{\citenamefont {Greve}\ \emph {et~al.}(2022)\citenamefont {Greve},
  \citenamefont {Luo}, \citenamefont {Wu},\ and\ \citenamefont
  {Thompson}}]{Greve2022}%
  \BibitemOpen
  \bibfield  {author} {\bibinfo {author} {\bibfnamefont {Graham~P}\
  \bibnamefont {Greve}}, \bibinfo {author} {\bibfnamefont {Chengyi}\
  \bibnamefont {Luo}}, \bibinfo {author} {\bibfnamefont {Baochen}\ \bibnamefont
  {Wu}}, \ and\ \bibinfo {author} {\bibfnamefont {James~K}\ \bibnamefont
  {Thompson}},\ }\bibfield  {title} {\enquote {\bibinfo {title}
  {{Entanglement-enhanced matter-wave interferometry in a high-finesse
  cavity}},}\ }\href {\doibase 10.1038/s41586-022-05197-9} {\bibfield
  {journal} {\bibinfo  {journal} {Nature}\ }\textbf {\bibinfo {volume} {610}},\
  \bibinfo {pages} {472--477} (\bibinfo {year} {2022})}\BibitemShut {NoStop}%
\bibitem [{\citenamefont {Gilmore}\ \emph {et~al.}(2021)\citenamefont
  {Gilmore}, \citenamefont {Affolter}, \citenamefont {Lewis-Swan},
  \citenamefont {Barberena}, \citenamefont {Jordan}, \citenamefont {Rey},\ and\
  \citenamefont {Bollinger}}]{Gilmore2021}%
  \BibitemOpen
  \bibfield  {author} {\bibinfo {author} {\bibfnamefont {Kevin~A.}\
  \bibnamefont {Gilmore}}, \bibinfo {author} {\bibfnamefont {Matthew}\
  \bibnamefont {Affolter}}, \bibinfo {author} {\bibfnamefont {Robert~J.}\
  \bibnamefont {Lewis-Swan}}, \bibinfo {author} {\bibfnamefont {Diego}\
  \bibnamefont {Barberena}}, \bibinfo {author} {\bibfnamefont {Elena}\
  \bibnamefont {Jordan}}, \bibinfo {author} {\bibfnamefont {Ana~Maria}\
  \bibnamefont {Rey}}, \ and\ \bibinfo {author} {\bibfnamefont {John~J.}\
  \bibnamefont {Bollinger}},\ }\bibfield  {title} {\enquote {\bibinfo {title}
  {{Quantum-enhanced sensing of displacements and electric fields with
  two-dimensional trapped-ion crystals}},}\ }\href {\doibase
  10.1126/science.abi5226} {\bibfield  {journal} {\bibinfo  {journal}
  {Science}\ }\textbf {\bibinfo {volume} {373}},\ \bibinfo {pages} {673--678}
  (\bibinfo {year} {2021})}\BibitemShut {NoStop}%
\bibitem [{\citenamefont {Gross}(2010)}]{Gross2012-2}%
  \BibitemOpen
  \bibfield  {author} {\bibinfo {author} {\bibfnamefont {Christian}\
  \bibnamefont {Gross}},\ }\bibfield  {title} {\enquote {\bibinfo {title} {Spin
  squeezing and non-linear atom interferometry with bose-einstein
  condensates},}\ }\href
  {http://books.google.de/books?id=lmtCqjtqg4EC&printsec=frontcover&dq=Spin+squeezing+and+non+linear+atom+interferometry+with+Bose+Einstein+condensates&cd=1&source=gbs_api}
  {\bibfield  {journal} {\bibinfo  {journal} {PhD dissertation}\ } (\bibinfo
  {year} {2010})}\BibitemShut {NoStop}%
\bibitem [{\citenamefont {Juli\'a-D\'{\i}az}\ \emph {et~al.}(2012)\citenamefont
  {Juli\'a-D\'{\i}az}, \citenamefont {Torrontegui}, \citenamefont {Martorell},
  \citenamefont {Muga},\ and\ \citenamefont {Polls}}]{PhysRevA.86.063623}%
  \BibitemOpen
  \bibfield  {author} {\bibinfo {author} {\bibfnamefont {B.}~\bibnamefont
  {Juli\'a-D\'{\i}az}}, \bibinfo {author} {\bibfnamefont {E.}~\bibnamefont
  {Torrontegui}}, \bibinfo {author} {\bibfnamefont {J.}~\bibnamefont
  {Martorell}}, \bibinfo {author} {\bibfnamefont {J.~G.}\ \bibnamefont {Muga}},
  \ and\ \bibinfo {author} {\bibfnamefont {A.}~\bibnamefont {Polls}},\
  }\bibfield  {title} {\enquote {\bibinfo {title} {Fast generation of
  spin-squeezed states in bosonic josephson junctions},}\ }\href {\doibase
  10.1103/PhysRevA.86.063623} {\bibfield  {journal} {\bibinfo  {journal} {Phys.
  Rev. A}\ }\textbf {\bibinfo {volume} {86}},\ \bibinfo {pages} {063623}
  (\bibinfo {year} {2012})}\BibitemShut {NoStop}%
\bibitem [{\citenamefont {Muessel}\ \emph {et~al.}(2015)\citenamefont
  {Muessel}, \citenamefont {Strobel}, \citenamefont {Linnemann}, \citenamefont
  {Zibold}, \citenamefont {Juli{\'a}-D{\'\i}az},\ and\ \citenamefont
  {Oberthaler}}]{Muessel2015}%
  \BibitemOpen
  \bibfield  {author} {\bibinfo {author} {\bibfnamefont {W.}~\bibnamefont
  {Muessel}}, \bibinfo {author} {\bibfnamefont {H.}~\bibnamefont {Strobel}},
  \bibinfo {author} {\bibfnamefont {D.}~\bibnamefont {Linnemann}}, \bibinfo
  {author} {\bibfnamefont {T.}~\bibnamefont {Zibold}}, \bibinfo {author}
  {\bibfnamefont {B.}~\bibnamefont {Juli{\'a}-D{\'\i}az}}, \ and\ \bibinfo
  {author} {\bibfnamefont {M.~K.}\ \bibnamefont {Oberthaler}},\ }\bibfield
  {title} {\enquote {\bibinfo {title} {Twist-and-turn spin squeezing in
  bose-einstein condensates},}\ }\href {\doibase 10.1103/PhysRevA.92.023603}
  {\bibfield  {journal} {\bibinfo  {journal} {Phys. Rev. A}\ }\textbf {\bibinfo
  {volume} {92}},\ \bibinfo {pages} {023603} (\bibinfo {year}
  {2015})}\BibitemShut {NoStop}%
\bibitem [{\citenamefont {Mirkhalaf}\ \emph {et~al.}(2018)\citenamefont
  {Mirkhalaf}, \citenamefont {Nolan},\ and\ \citenamefont
  {Haine}}]{Mirkhalaf2018}%
  \BibitemOpen
  \bibfield  {author} {\bibinfo {author} {\bibfnamefont {Safoura~S.}\
  \bibnamefont {Mirkhalaf}}, \bibinfo {author} {\bibfnamefont {Samuel~P.}\
  \bibnamefont {Nolan}}, \ and\ \bibinfo {author} {\bibfnamefont {Simon~A.}\
  \bibnamefont {Haine}},\ }\bibfield  {title} {\enquote {\bibinfo {title}
  {Robustifying twist-and-turn entanglement with interaction-based readout},}\
  }\href {\doibase 10.1103/PhysRevA.97.053618} {\bibfield  {journal} {\bibinfo
  {journal} {Phys. Rev. A}\ }\textbf {\bibinfo {volume} {97}},\ \bibinfo
  {pages} {053618} (\bibinfo {year} {2018})}\BibitemShut {NoStop}%
\bibitem [{\citenamefont {Sorelli}\ \emph {et~al.}(2019)\citenamefont
  {Sorelli}, \citenamefont {Gessner}, \citenamefont {Smerzi},\ and\
  \citenamefont {Pezz{\`e}}}]{Sorelli2019}%
  \BibitemOpen
  \bibfield  {author} {\bibinfo {author} {\bibfnamefont {Giacomo}\ \bibnamefont
  {Sorelli}}, \bibinfo {author} {\bibfnamefont {Manuel}\ \bibnamefont
  {Gessner}}, \bibinfo {author} {\bibfnamefont {Augusto}\ \bibnamefont
  {Smerzi}}, \ and\ \bibinfo {author} {\bibfnamefont {Luca}\ \bibnamefont
  {Pezz{\`e}}},\ }\bibfield  {title} {\enquote {\bibinfo {title} {Fast and
  optimal generation of entanglement in bosonic josephson junctions},}\ }\href
  {\doibase 10.1103/PhysRevA.99.022329} {\bibfield  {journal} {\bibinfo
  {journal} {Phys. Rev. A}\ }\textbf {\bibinfo {volume} {99}},\ \bibinfo
  {pages} {022329} (\bibinfo {year} {2019})}\BibitemShut {NoStop}%
\end{thebibliography}
\end{document}